\documentclass[aps,prd,twocolumn,groupedaddress,showpacs,preprintnumbers,floatfix]{revtex4}
\usepackage{amssymb}
\usepackage{graphicx}
\usepackage{epsfig}
\usepackage{amsmath}
\newcommand{\beq}{\begin{equation}} 
\newcommand{\eeq}{\end{equation}}   
\newcommand{\bea}{\begin{eqnarray}} 
\newcommand{\eea}{\end{eqnarray}}
\def\Li2{\hbox{Li}_2}

\begin{document}

\title{Four-pion production in $\tau$ decays and $e^+e^-$ annihilation:
  an update}
\thanks{Work 
supported in part by BMBF grant 05HT4VKA/3,
EU 6th Framework Program under contract MRTN-CT-2006-035482 
(FLAVIAnet), TARI project RII3-CT-2004-506078,
Polish State Committee for Scientific Research (KBN)
under contract 1 P03B 003 28}

\author{Henryk Czy\.z}
\affiliation{Institute of Physics, University of Silesia,
PL-40007 Katowice, Poland.}

\author{Johann H. K\"uhn}

\affiliation{Institut f\"ur Theoretische Teilchenphysik,
Universit\"at Karlsruhe, D-76128 Karlsruhe, Germany.}

\author{Agnieszka Wapienik}
\affiliation{Institute of Physics, University of Silesia,
PL-40007 Katowice, Poland}

\date{\today}

\begin{abstract}
  An improved description of four-pion production in electron-positron
annihilation and in $\tau$-lepton decays is presented. The model amplitude
is fitted to recent data from BaBar which cover a wide energy range and
which were obtained exploiting the radiative return. Predicting
$\tau$-decay distributions from $e^+e^-$ data and comparing these
predictions with ALEPH and CLEO results, the validity of isospin symmetry
is confirmed within the present experimental errors. A good description of
two- and three-pion sub-distributions is obtained. Special emphasis is put on
the predictions for $\omega\pi\,(\to\pi^+\pi^-\pi^0)$ in $e^+e^-$
annihilation and in $\tau$ decay. The model amplitude is implemented in
the Monte Carlo generator PHOKHARA.
\end{abstract}


\maketitle
\newcommand{\Eq}[1]{Eq.(\ref{#1})} 

\section{Introduction}

The production of four pions in $\tau$ decays and $e^+e^-$ annihilation
 has received considerable attention, both from the theoretical \cite{Fischer:1979fh,Decker:1987mn,Decker:1994af,Kuhn:1998rh,Czyz:2000wh,Ecker:2002cw,Davier:2002dy,Bondar:2002mw,Davier:2005xq}
and the experimental side  \cite{Akhmetshin:1998df,Edwards:1999fj,Akhmetshin:1999ty,Achasov:2003bv,Akhmetshin:2004dy,Schael:2005am,Aubert:2005eg,bb1}. Relating the cross sections and rates for the four charge
combinations ($\pi^+\pi^- 2\pi^0$, $2\pi^+2\pi^-$, $2\pi^-\pi^+ \pi^0$
and $\pi^- 3\pi^0$) gives important hints on the validity of isospin symmetry
and the size of the isospin breaking terms. The dependence of the rates
and the cross sections on $\sqrt{Q^2}$, the invariant mass of the four-pion
 system, and the investigation of differential distributions, e.g. of
the two- and/or three-pion masses, gives information on the resonance
structure of the amplitude. In the low $Q^2$ region, predictions
based on the chiral Lagrangian can be tested which, however, must be 
complemented by resonance physics in order to properly describe
the rates in the dominant region between 1 and 3 GeV. The $e^+e^-$ cross
section is, furthermore, important to evaluate the hadronic vacuum
polarization which in turn is essential for the precise prediction
of the muon anomalous magnetic moment and the running of the electromagnetic
coupling \cite{fred,MelnikovVainshtein}.

 From the experimental side precise $\tau$ data have been obtained
 by ALEPH \cite{Schael:2005am} and by CLEO \cite{Edwards:1999fj}
 collaborations, which, however, are naturally restricted to
 $\sqrt{Q^2}$ below 1.77 GeV. The $e^+e^-$ cross section has been 
 measured by CMD2 \cite{Akhmetshin:1998df,Akhmetshin:1999ty,Akhmetshin:2004dy}  and SND \cite{Achasov:2003bv} (older  data
 are far less accurate and will not used in this paper)
 and, more recently, by BaBar \cite{Aubert:2005eg,bb1} through the
 method of radiative return which covers energies up to 4.5 GeV.
 This method, which was proposed in \cite{Binner:1999bt,Zerwas,Czyz:2000wh} 
 allows to use the large luminosity at $B$-factories for a measurement
 of the $e^+e^-$ cross  section in the region of interest.
 
 From the theory side the first evaluation based on chiral perturbation
 theory has been performed by Fi\-scher, Wagner and Wess \cite{Fischer:1979fh}
 and applied to $\tau$ decays. Subsequently this ansatz was
 extended \cite{Decker:1994af} to include $\rho$, $a_1$ and $f_0$ resonances,
  which are clearly visible in sub-distributions. In addition the $\omega\pi$
 mode was introduced, again predicted
 from the chiral anomaly \cite{Ecker:2002cw,Decker:1987mn}. Later this ansatz,
 slightly modified, was implemented in the generator EVA \cite{Binner:1999bt} to simulate
 $4\pi$ production in the radiative return \cite{Czyz:2000wh}.
 As stated above, the low $Q^2$ region should be best suited for a description
 based on chiral Lagrangians. Combining one-loop chiral corrections at low
 $Q^2$ with resonance enhancements at intermediate energies, precise predictions
 have been obtained in \cite{Ecker:2002cw} which will be discussed below.

 In view of these recent theoretical and experimental developments, together
 with need for an optimal implementation of the $4\pi$ mode into the 
 Monte Carlo event generator PHOKHARA \cite{Rodrigo:2001kf,Czyz:2002np,Czyz:PH03,Nowak,Czyz:PH04,Czyz:2004nq,Czyz:2005as,Czyz:2007wi},
  an improved ansatz for the corresponding
 hadronic amplitude has been developed. The ansatz is largely based on
 \cite{Decker:1994af,Czyz:2000wh} and \cite{Czyz:2005as} (concerning the
 $\omega$ part) with model parameters fitted to the recent BaBar results.
 In order to accommodate the $\rho^+\rho^-$ signal observed in \cite{bb1}
 we include a contribution, which is modeled to
 mimic a $SU(2)$ gauge theory with the $\rho$- meson 
 (and its radial excitations) as gauge boson(s).

  Our paper is organized as follows: To facilitate the subsequent discussion,
 in Section \ref{definitions} the basic definitions are introduced and
 the (well known) isospin relations between the amplitudes and the rates
 of the four channels are collected. The validity of these relations
 is investigated in Section \ref{expsituation}, using data from $e^+e^-$
 annihilation to predict the corresponding, experimentally measured
  distributions for $\tau$ decays.
 The ingredients for the ansatz for the matrix element of the hadronic current
 are discussed in Section \ref{modeldescription}. 
 The comparison of this ansatz with $e^+e^-$ data and the fit of its parameters
 is presented in Section \ref{fittingthedata} together with the comparison
 between the model and data for a variety of distributions. The implications 
 of the model for $\tau$ decays is discussed in Section \ref{taudecays},
 the implementation into the generator PHOKHARA and related technical
 tests in Section \ref{PHOKHARA}. A brief summary and our conclusions are
 given in Section \ref{conclusions}. A detailed description of our model
 with the complete list of parameters can be found in the Appendix.

\section{General properties of the four pion electromagnetic current}
\label{definitions}
  
 General properties of the four pion electromagnetic current were
 investigated in \cite{Kuhn:1998rh}, where it was shown
 that assuming isospin symmetry just one function $J_{\mu}$
 describes all four matrix elements. 
 We use the same letter $J$ for the operator and its matrix element.
 For convenience we recall
  the definitions and notation introduced in
  \cite{Kuhn:1998rh,Czyz:2000wh}.

Ignoring the issues of isospin breaking and radiative corrections,
the electromagnetic current can be decomposed into an isospin singlet piece
and a part transforming like the third component of an isospin triplet:

\bea
 J^{em} = \frac{1}{\sqrt{2}}\ \  J^3 + \frac{1}{3\sqrt{2}}\ \  J^{{\rm I}=0}
\eea

\noindent
 whereas the charged current generating \(\tau\) decays is given by

\bea
J^- = \frac{1}{\sqrt{2}} \ \ (J^1-i \ J^2) \ \ .
\eea

Final states with an even number of pions are produced through the isospin
 one part only
\bea
\langle \pi^+ \pi^- \pi_1^0 \pi_2^0 | J^3_{\mu} | 0 \rangle &=& 
J_{\mu}(p_1,p_2,p^+,p^-) \nonumber \\\nonumber \\
\langle \pi^+_1 \pi^+_2 \pi^-_1 \pi^-_2 | J^3_{\mu} | 0 \rangle &=&\nonumber \\
J_{\mu}(p_2^+,p_2^-,p_1^+,p_1^-) \kern-20pt&&+
J_{\mu}(p_1^+,p_2^-,p_2^+,p_1^-) \nonumber \\
+J_{\mu}(p_2^+,p_1^-,p_1^+,p_2^-)\kern-20pt&&+
 J_{\mu}(p_1^+,p_1^-,p_2^+,p_2^-)\nonumber \\\nonumber \\
\langle \pi^- \pi^0_1 \pi^0_2 \pi^0_3 | J^{-}_{\mu} | 0 \rangle &=& 
\nonumber \\
&&\kern-100pt J_{\mu}(p_2,p_3,p^-,p_1) +
J_{\mu}(p_1,p_3,p^-,p_2)+J_{\mu}(p_1,p_2,p^-,p_3)
\nonumber \\\nonumber \\
\langle \pi^-_1 \pi^-_2 \pi^+ \pi^0 | J^{-}_{\mu} | 0 \rangle &=& \nonumber \\
J_{\mu}(p^+,p_2,p_1,p^0)\kern-20pt&&+
J_{\mu}(p^+,p_1,p_2,p^0)
 \label{rel} \ .
\eea
The function 
 \(J_{\mu} \equiv J_\mu(q_1,q_2,q_3,q_4) \) is symmetric (anti-symmetric)
 with respect to the interchange of \(q_1\) and \(q_2\) (\(q_3\) and \(q_4\)).

 In \cite{Ecker:2002cw} it was shown that also the matrix element
 $\langle \pi^-_1 \pi^-_2 \pi^+ \pi^0 | J^{-}_{\mu} | 0 \rangle$
 can be used as an independent function, through which the other ones can
 be expressed. If one denotes

\bea
&&\kern-20pt\langle \pi^- \pi^0_1 \pi^0_2 \pi^0_3 | J^{-}_{\mu} | 0 \rangle = 
 F_{\mu}(p_1,p_2,p^-,p_3)\ , 
 \label{rel2} 
\eea
one gets relation
 \bea
&&\kern-20pt\langle \pi^+_1 \pi^+_2 \pi^-_1 \pi^-_2 | J^3_{\mu} | 0 \rangle =\nonumber \\
 &&\kern+15pt  F_{\mu}(p_2^+,p_2^-,p_1^+,p_1^-)+F_{\mu}(p_2^+,p_1^+,p_2^-,p_1^-)
 \label{rel1} \ ,
\eea
 from which it is clear that the matrix element
  $\langle \pi^+_1 \pi^+_2 \pi^-_1 \pi^-_2 | J^3_{\mu} | 0 \rangle$
   can
 be expressed by the matrix element
  $\langle \pi^- \pi^0_1 \pi^0_2 \pi^0_3 | J^{-}_{\mu} | 0 \rangle$.
 The opposite is also true and we have proved it using the method
 developed in \cite{Ecker:2002cw} to express 
 $\langle \pi^+ \pi^- \pi_1^0 \pi_2^0 | J^3_{\mu} | 0 \rangle$ by 
 $\langle \pi^-_1 \pi^-_2 \pi^+ \pi^0 | J^{-}_{\mu} | 0 \rangle$. However,
 as in both cases the obtained inverse relations are far from being as
  elegant as the ones of Eq. (\ref{rel}) and Eq. (\ref{rel1}),
   the result is not presented here.

The currents defined in Eq.(\ref{rel}) contain the complete information
 about the hadronic cross section through 

\begin{eqnarray}
 &&\kern-60pt \int \ J^{em}_\mu (J^{em}_\nu)^*  \ \ d\bar\Phi_n(Q;q_1,\dots,q_n)
 \nonumber \\  &=&  
 \frac{1}{6\pi} \left(Q_{\mu}Q_{\nu}-g_{\mu\nu}Q^2\right) \ R(Q^2)
\label{rr}
\end{eqnarray}
 
\noindent
where \(R(Q^2)\) is equal to \(\sigma(e^+e^-\rightarrow hadrons)(Q^2)/\sigma_{point}\),
with $\sigma_{point}=4\pi\alpha^2/(3Q^2)$,
 and \(d\bar\Phi_n(Q;q_1,\dots,q_n)\) denotes the \(n\) body phase space
 with all statistical factors included.

The amplitude describing the \(\tau\) decay into an arbitrary
 number of hadrons plus a neutrino (excluding radiative corrections)
   is given by

\begin{eqnarray}
 {\cal M}_{\tau} = \ \ \frac{G_F}{\sqrt{2}} \ V_{ud} \ \
 \bar v\left(p_{\nu}\right) \gamma^{\alpha}\left(1-\gamma_5\right)
 u\left(p_{\tau}\right) \ J_{\alpha}^{-} \ ,
\end{eqnarray}
with 

\begin{equation}
 J^{-}_{\alpha} \ \equiv \ J^{-}_{\alpha}\left(q_1,...,q_n\right) \ \equiv \ 
 \langle h(q_1),...,h(q_n)|J^{-}_{\alpha}\left(0\right)|0\rangle 
\end{equation}
and \(J^{-}_{\alpha}(0) \equiv \bar d \gamma_\alpha  u\) at the quark level,
where we
restrict our considerations to the Cabbibo allowed vector part of the
hadronic current.

 The differential $\tau$ decay rates are given by

\begin{eqnarray}
&&\kern-30pt\frac{d\Gamma_{\tau\to\nu+hadrons}}{dQ^2} \nonumber \\
 &&\kern-30pt= 2 \ \Gamma_e \frac{|V_{ud}|^2S_{EW}}{m_{\tau}^2}
 \left(1-\frac{Q^2}{m_{\tau}^2}\right)^2 
 \left(1+2\frac{Q^2}{m_{\tau}^2}\right) R^{\tau}\left(Q^2\right) \ ,
\label{diff}
\end{eqnarray}
with 

\begin{eqnarray}
 &&\kern-30pt
 \int \ J_\mu^{-} J^{-*}_\nu  \ \ d\bar\Phi_n(Q;q_1,\dots,q_n) = \nonumber \\
 &&\frac{1}{3\pi} \left(Q_{\mu}Q_{\nu}-g_{\mu\nu}Q^2\right) \ R^{\tau}(Q^2) \ 
\label{rrt}
\end{eqnarray}
\noindent
 and $\Gamma_e= G_F^2m_\tau^5/(192\pi^3) $.
Note the relative factor of 2 between the definitions in \Eq{rr} and 
 \Eq{rrt}. We have included here also electroweak correction factor
 $S_{EW}$ (we use $S_{EW}=1.0198$ \cite{Davier:2002dy}) to account 
 for standard electroweak corrections.

  The function $R^{\tau}$ is related to the spectral function defined by 
  CLEO \cite{Edwards:1999fj} through $R^{\tau}(--+0) = 3\pi V^{3\pi\pi^0}$
  and through  $R^{\tau} =  3v_1$ to the vector spectral functions defined
  by ALEPH \cite{Schael:2005am}. In this paper we will use normalization
  of the spectral functions chosen by ALEPH.

   The four pion spectral functions and the cross sections can
  be expressed as linear combinations of two integrals

\begin{widetext}
   \bea
 A &=& -\frac{2\pi}{Q^2}
  \int \ J^\mu\left(q_1,q_2,q_3,q_4\right)
             J^{*}_\mu \left(q_1,q_2,q_3,q_4\right)
    \ d\Phi_4(Q;q_1,\dots,q_4)\nonumber \\
 B &=& -\frac{4\pi}{Q^2}
  \int 
 \ {\rm Re}\left(J^\mu\left(q_1,q_2,q_3,q_4\right)
             J^{*}_\mu \left(q_1,q_3,q_2,q_4\right) \right)
    \ d\Phi_4(Q;q_1,\dots,q_4) \ .
\eea
  The relations read
\bea
 R\left(+\ - \ 0 \ 0\right) &=& \frac{1}{2} \ A
 {\kern+33pt} \ \ \ ; \ \ \ 
 R^{\tau}\left(-\ - \ + \ 0\right) \ = \  A + \frac{1}{2} \ B 
  \nonumber \\
 R\left(+\ + \ - \ -\right) &=&  A \ +  \ B  
 \ \ \ \ \ \ \ \  ; \ \ \ 
   R^{\tau}\left(-\ 0 \ 0 \ 0\right) {\kern+9pt}
 \ = \ \frac{1}{2}\left( A \ +  \ B\right)  \ .
\label{isorel}
\eea 
  The additional contribution to
  $R\left(+\ + \ - \ -\right) =  A \ +  \ B+  \ C$ of the form 

 \bea
C = -\frac{2\pi}{Q^2}
  \int 
 \ {\rm Re}\left(J^\mu\left(q_1,q_2,q_3,q_4\right)
             J^{*}_\mu \left(q_3,q_4,q_1,q_2\right) \right)
    \ d\Phi_4(Q;q_1,\dots,q_4) \ ,
\eea
\end{widetext}
 vanishes for any symmetric phase space configuration.
\noindent
 Eqs.\ref{isorel} 
 correspond to  the familiar relations between \(\tau\) decay rates and
 \(e^+e^-\) annihilation cross sections:

\bea
 R^{\tau}\left(-\ 0 \ 0 \ 0\right) &=&  \frac{1}{2}\  R\left(+\ + \ - \ -\right)
\nonumber \\
 R^{\tau}\left(-\ - \ + \ 0\right) 
 &=& \frac{1}{2} \ R\left(+\ + \ - \ -\right) \ \nonumber \\
 &+& \ R\left(+\ - \ 0 \ 0\right) \ .
\label{CVC}
\eea
\section{Isospin symmetry - experimental situation}\label{expsituation}
\begin{figure}[ht]
\includegraphics[width=8.5cm,height=6cm]{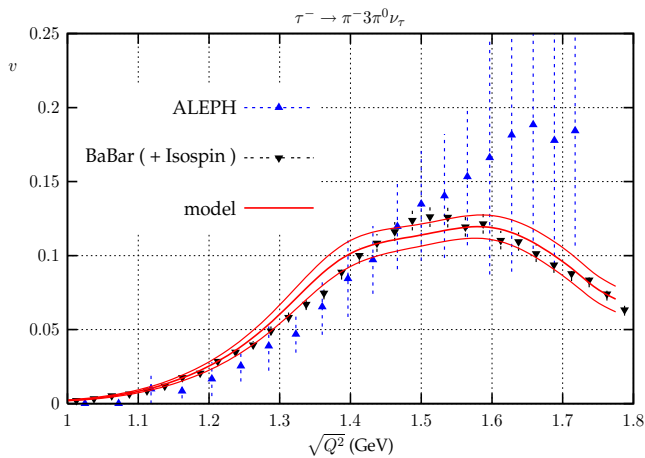}
\caption{(color online).  
 The  spectral function of the $\tau^- \to \nu 3\pi^0 \pi^-$
 decay mode. ALEPH data \cite{Schael:2005am} versus predictions
 from BaBar data \cite{Aubert:2005eg,bb1} and the model predictions.
}
\label{spec1}
\end{figure}

\begin{figure}[ht]
\includegraphics[width=8.5cm,height=6cm]{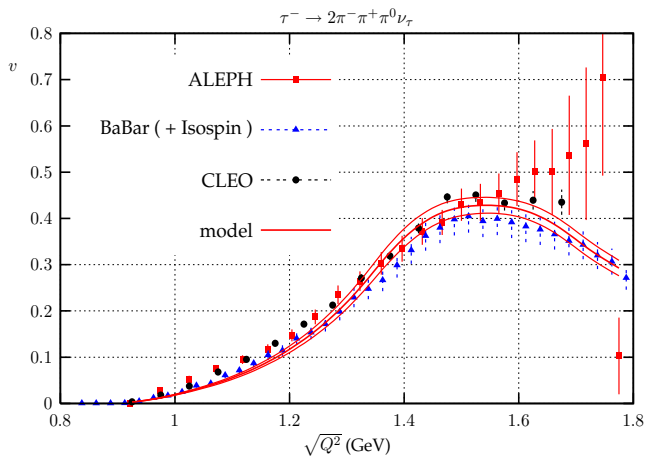}
\caption{(color online).  
 The  spectral function of the $\tau^- \to \nu 2\pi^- \pi^+\pi^0$
 decay mode. ALEPH \cite{Schael:2005am} and CLEO \cite{Edwards:1999fj}
 data versus predictions
 from BaBar data \cite{Aubert:2005eg,bb1} and the model predictions.
}
\label{spec2}
\end{figure}

 In this section we would like to address the question, if
 present experiments require
 inclusion of isospin violating effects in the model. 
 Combining the results from BaBar \cite{Aubert:2005eg}
 on  $\sigma(e^+e^-\to 2\pi^+2\pi^-)$
  with their preliminary results on $\sigma(e^+e^-\to 2\pi^0\pi^+\pi^-)$
 \cite{bb1} and  using Eqs.(\ref{CVC}),
 one obtains predictions for the $\tau$ spectral functions.
  These can be compared with 
 ALEPH \cite{Schael:2005am} and
  CLEO \cite{Edwards:1999fj} data (compare also \cite{bb1}).
   As shown in Fig. \ref{spec1} and Fig. \ref{spec2},
  $\tau$- and $e^+e^-$- data are in good agreement within the
  errors, even if one observes  systematical shifts. 
  However, these shifts are well within the 5\% systematic error of CLEO and the 6\% 
  ($2\pi^-\pi^+\pi^0$) and 10\% ($\pi^- 3\pi^0$) errors
 for ALEPH spectral functions (ALEPH does not give separately
  the systematic error) as well as the 5\% to 12\% systematic error for BaBar 
 $\sigma(e^+e^-\to 2\pi^+2\pi^-)$. For the preliminary BaBar data \cite{bb1}
  on $\sigma(e^+e^-\to 2\pi^0\pi^+\pi^-)$ a 10\% systematic error is assumed.
   Truly isospin breaking effects are expected to occur 
  at the percent level due to the $\pi^\pm-\pi^0$ mass difference alone
 \cite{Czyz:2000wh}.  

\begin{figure}[ht]
\includegraphics[width=8.5cm,height=6cm]{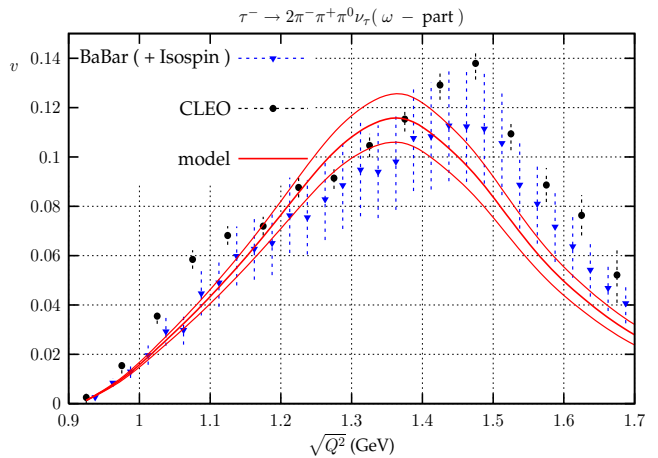}
\caption{(color online).  
 The  omega part of the 
 spectral function of the $\tau^- \to \nu 2\pi^- \pi^+\pi^0$
 decay mode. CLEO \cite{Edwards:1999fj}
 data versus predictions
 based on preliminary BaBar data \cite{bb1} and the model predictions.
}
\label{spec3}
\end{figure}

  From the cross section $\sigma(e^+e^-\to 2\pi^0\pi^+\pi^-)$ and the relative
  contributions of the $\omega\pi$ final state as given in \cite{bb1}
  (since the errors were not specified there, we attribute
   20\% error to the spectrum) 
  one can infer the $\omega$ contribution to 
  $\sigma(e^+e^-\to 2\pi^0\pi^+\pi^-)$. Based on this result one
  can predict  the omega part of the 
 $\tau^- \to \nu 2\pi^- \pi^+\pi^0$ spectral function and compare it
  with the CLEO result. Satisfactory agreement is observed
  in  Fig. \ref{spec3}.

  From the comparisons of experimental data we conclude
  that no isospin symmetry violation is observed within the present accuracy.
   Thus the model we propose to describe the data is based
  on isospin symmetry. However, effects from the pion mass difference
   in the phase space are included.

\section{The model of the four pion electromagnetic current}
 \label{modeldescription}

 There are many motivations why the model adopted in \cite{Czyz:2000wh}
  should be updated. First of all
 new and more accurate data are available. The CLEO data on tau decays
 \cite{Edwards:1999fj},
 which were not used in \cite{Czyz:2000wh}, the  tau spectral
 functions from ALEPH \cite{Schael:2005am} and the measurement
 of the cross section of the reaction $e^+e^-\to 2\pi^+2\pi^-$ 
 via the radiative return method by BaBar \cite{Aubert:2005eg} provide us
 with the opportunity for a substantial improvement of the model implemented
 in the event generator PHOKHARA 
 \cite{Rodrigo:2001kf,Czyz:2002np,Czyz:PH03,Czyz:PH04,Czyz:2004nq,Nowak,Czyz:2005as,Czyz:2007wi}.
  The omega part of the current, which in \cite{Decker:1994af,Czyz:2000wh}
 was implemented without
  structure, is now known much better from phenomenological studies 
  \cite{Czyz:2005as}. The new preliminary data from BaBar \cite{bb1} 
 on the reaction $e^+e^-\to 2\pi^0\pi^+\pi^-$ 
 also show richer structure than implemented in
 \cite{Czyz:2000wh}.
 All this was taken into account in constructing the model 
  presented in this paper.

\begin{figure}[ht]
\begin{center}
\epsfig{file=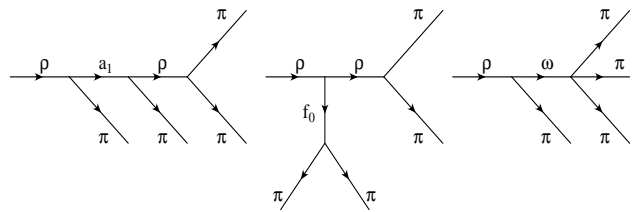,width=1.cm,height=0.8cm}
\end{center}
\phantom{}
\vskip 1.5cm
\caption{Diagrams contributing to the hadronic current in \cite{Czyz:2000wh}.
}
\label{origc}
\end{figure}

\begin{figure}[ht]
\includegraphics[width=6.5cm,height=4cm]{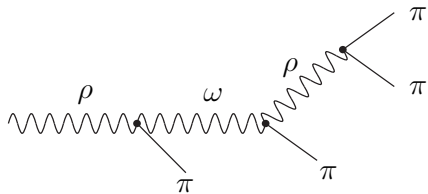}
\caption{ 
 The new contributions from the omega part of the current.
}
\label{omega}
\end{figure}

 The amplitude used in \cite{Czyz:2000wh} is  schematically
 depicted in Fig. \ref{origc}. In the contributions from the first two 
  diagrams, which proceed through the intermediate resonances
 $\rho\to a_1 \pi$ and $\rho \to f_0\rho$ respectively (where $\rho$
 stands for $\rho(770)$ and its radial excitations), 
  only the parameters
 of the current were adopted to the improved data and
 a new $\rho$ resonance ($\rho(2040)$) was
 added (necessary to fit the BaBar \cite{Aubert:2005eg} data). The
 contribution from $\omega$, where previously the substructure of the omega
 decay was not taken into account, is now modeled using information
 from \cite{Czyz:2005as}. Schematically, the new $\omega$ amplitude
  is depicted in Fig.\ref{omega}.

\begin{figure}[ht]
\includegraphics[width=4.1cm,height=3cm]{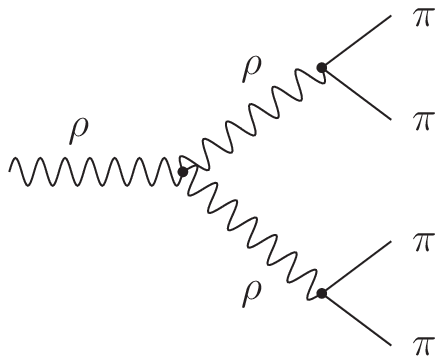}
\includegraphics[width=4.1cm,height=3cm]{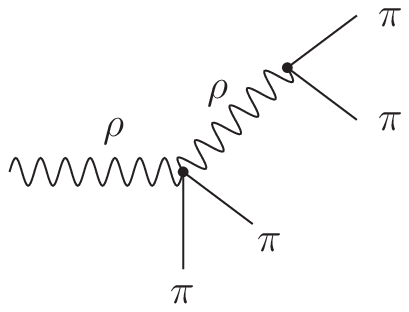}
\caption{ 
 The new contributions from $\rho$ mesons.
}
\label{rhorho}
\end{figure}

  BaBar has,  furthermore, observed \cite{bb1}  a strong 
  $\rho^+\rho^-$ contribution.
 Thus a new part containing the $\rho\to\rho \rho$ contribution
 has been added,
 treating the $\rho$ particles like $SU(2)$ gauge bosons. 
 The contributions to the amplitude are depicted in Fig. \ref{rhorho}.
 For more general frameworks, where such terms are present see
   \cite{Harada:2003jx}
 (and references therein).
 The SU(2) symmetric Lagrangian describing $\rho -$ pair production
 reads 

\bea
 {\cal L}_\rho &=& \frac{1}{4} {\overset{\rightarrow}{ F}}_{\mu\nu}\cdot
{\overset{\rightarrow}{ F}}^{\mu\nu} 
  + \frac{1}{2}\overset{\longrightarrow}{\left(D^\mu{\phi}\right)}\cdot
 \overset{\longrightarrow}{\left( D_\mu{{\phi}}\right)}  \nonumber \\
 &+&  \frac{1}{2} m_\pi^2
         \overset{\rightarrow}{\phi}\cdot\overset{\rightarrow}{\phi}
 +  \frac{1}{2}  m_\rho^2\overset{\rightarrow}{\rho}_\mu
     \cdot\overset{\rightarrow}{\rho}^\mu
\ ,
\label{lagrangian}
\eea
where  

\bea
 &&\overset{\rightarrow}{ \phi} = \left(
    \begin{array}{c}
     \frac{1}{\sqrt{2}}\left(\pi^++\pi^-\right) \\
      \frac{{\rm i}}{\sqrt{2}}\left(\pi^+-\pi^-\right)\\
       \pi^0
    \end{array}
    \right) \ , \ 
 \overset{\rightarrow}{ \rho}_\mu = \left(
    \begin{array}{c}
     \frac{1}{\sqrt{2}}\left(\rho^++\rho^-\right) \\
      \frac{{\rm i}}{\sqrt{2}}\left(\rho^+-\rho^-\right)\\
       \rho^0
    \end{array}
    \right)_\mu \nonumber \\
&&{ \overset{\rightarrow}{F}}_{\mu\nu} =
 \partial_\mu{\overset{\rightarrow}{ \rho}}_\nu 
- \partial_\nu{\overset{\rightarrow}{ \rho}}_\mu
 - g {\overset{\rightarrow}{ \rho}}_\mu \times\overset{\rightarrow}{ \rho}{}_\nu 
\eea 

and 
   \bea
 \overset{\longrightarrow} {D_\mu{{\phi}}}   = 
   \partial_\mu \overset{\rightarrow}{ \phi}
  + g \left({\overset{\rightarrow}{ \rho}}_\mu \times 
 \overset{\rightarrow}{ \phi}\right) \ .
   \eea

  The only free parameter, the coupling constant $g$ ($g=g_{\rho\pi\pi}$),
  can be extracted 
  from $\rho\to\pi\pi$ decay.  However, as it stands, the model 
  leads to a wrong high energy behavior
  of the cross section, falling less rapidly then the data.
  This problem can be cured by adding $\rho'$ contributions
  and allowing for trilinear couplings between $\rho$ and $\rho'$.
  It was also necessary to relax the fixed coupling $g$ to fit the data.
  The detailed description can be found in   the Appendix.
   The model can be further refined,
   when more experimental information is available. 

  The behavior of the four pion amplitude
 in the low $Q^2$ region has also been studied \cite{Ecker:2002cw}
  in the framework of chiral resonance theory, including terms up
 to $O(p^4)$ \cite{Ecker:1988te}. The implementation of  resonances
 and their parameters differs from the choice in this paper.
 The results of the two models are compared to the data 
 in Figs. \ref{chargedEcker} and \ref{neutralEcker}.


\section{Fit of the current parameters to the experimental data}
\label{fittingthedata}

 To separate the  well measured $\omega$
  contribution, from the rest,
  we fitted the parameters of the model
   to the
  $\omega$ part of the cross section of the reaction
  $e^+e^-\to 2\pi^0\pi^+\pi^-$ extracted from preliminary BaBar 
  data \cite{bb1}. Furthermore, we fitted the model parameters
   to the cross sections of the reactions $e^+e^-\to 2\pi^0\pi^+\pi^-$
  and $e^+e^-\to 2\pi^+2\pi^-$ measured by BaBar \cite{bb1,Aubert:2005eg}.

\begin{figure}[ht]
\includegraphics[width=8.5cm,height=7cm]
 {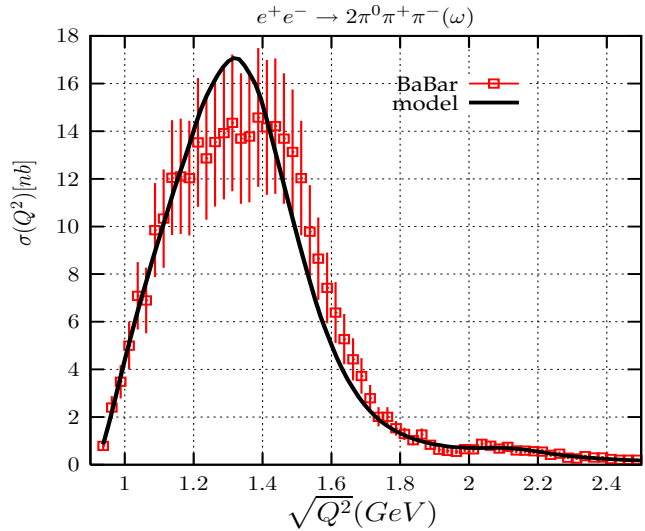}
\caption{(color online).  
 Fit to the  $\omega-$ part of the  cross
section for $e^+e^-\to 2\pi^0\pi^+\pi^-$ \cite{bb1}
  (20\% systematical error for the preliminary BaBar data was assumed).
}
\label{omegafit} 
\end{figure}

\begin{figure}[ht]
\includegraphics[width=8.5cm,height=7cm]
 {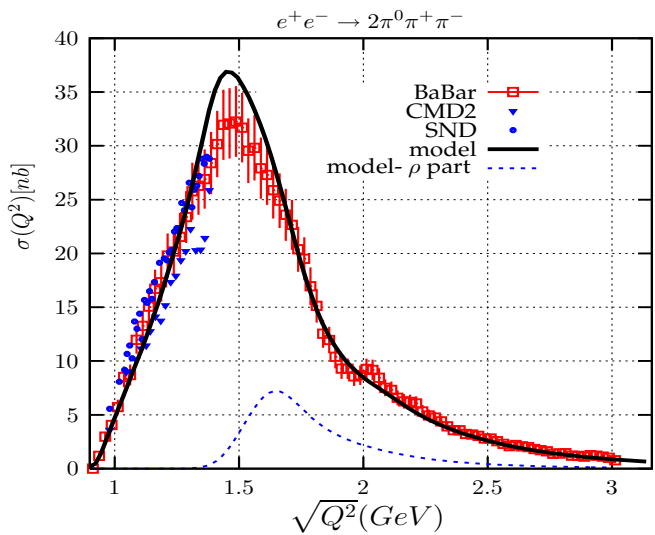}
\caption{(color online).  
  Fit to the  data  
  for  $\sigma(e^+e^-\to 2\pi^0\pi^+\pi^-)$, taken from \cite{bb1}
 (10\% systematical error was added to the statistical error). 
   For comparison also
CMD2 \cite{Akhmetshin:1998df}
 and SND \cite{Achasov:2003bv} data, which are consistent with
 BaBar data, are shown (without their 10\%-20\% error-bars).
Contributions from  $\rho$ part of the current
 (\Eq{rhorhocurrent}) to the cross section (see text for definition)
 are also shown.}
\label{neutralfit} 
\end{figure}

\begin{figure}[ht]
\includegraphics[width=8.5cm,height=7cm]
 {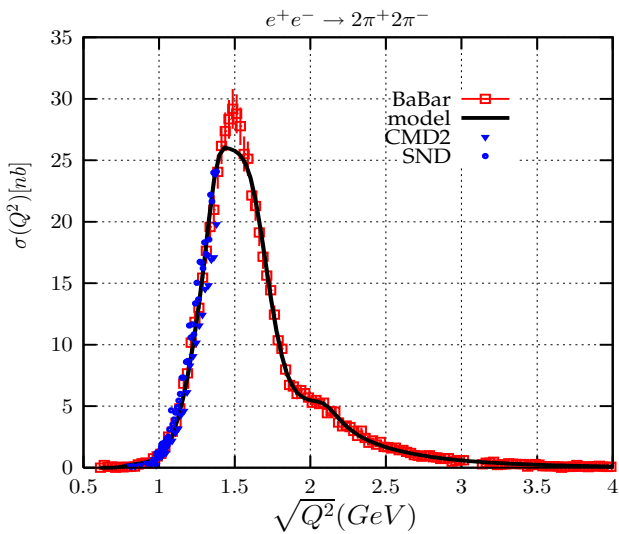}
\caption{(color online).  
  Fit to the   data
  for  $\sigma(e^+e^-\to 2\pi^+2\pi^-)$, taken from \cite{Aubert:2005eg}.
 For comparison also
CMD2 \cite{Akhmetshin:1998df,Akhmetshin:1999ty,Akhmetshin:2004dy}
 and SND \cite{Achasov:2003bv} data, which are consistent with
 BaBar data, are shown (without their 7\%-20\% error-bars).
}
\label{chargedfit} 
\end{figure}

\begin{figure}[ht]
\includegraphics[width=8.5cm,height=7cm]{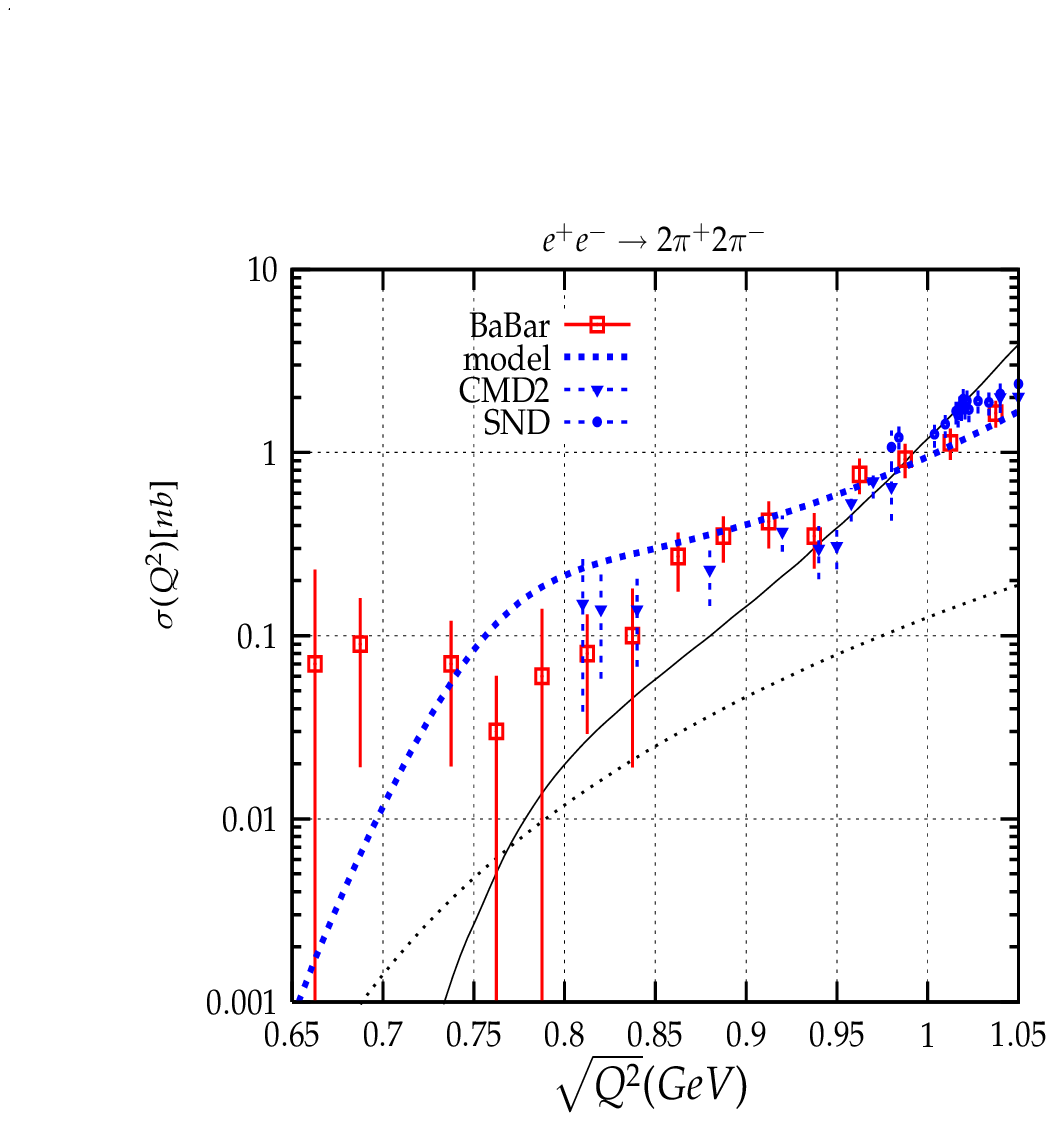}
\caption{(color online).  
 Comparison of our model with the chiral Lagrangian \cite{Ecker:2002cw}
 predictions (solid and dotted lines) in low energy region
 and experimental data 
 \cite{Akhmetshin:1999ty,Achasov:2003bv,Akhmetshin:2004dy,Aubert:2005eg}.
}
\label{chargedEcker}
\end{figure}
\begin{figure}[ht]
\includegraphics[width=8.5cm,height=7cm]{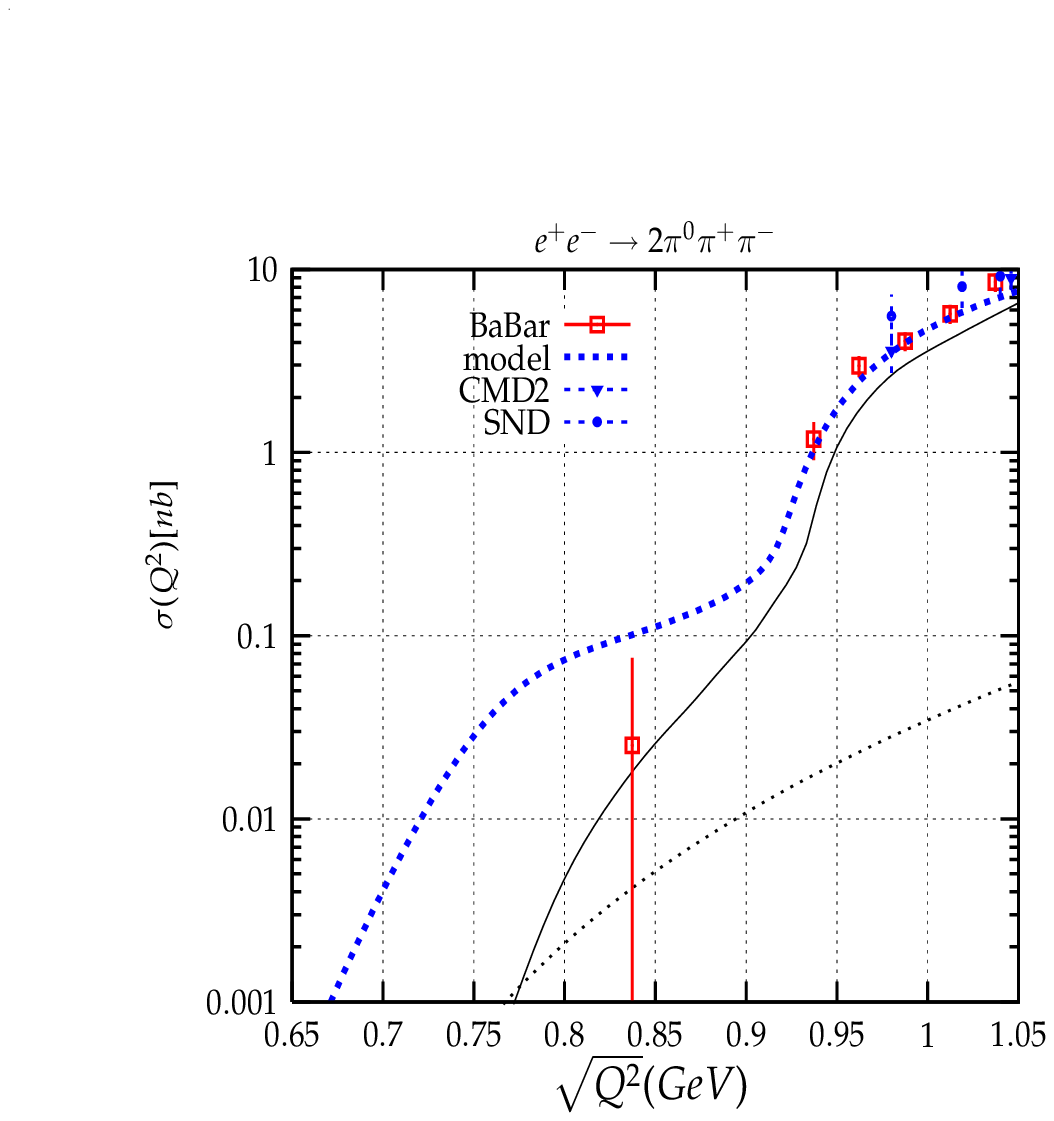}
\caption{(color online).  
 Comparison of our model with the chiral Lagrangian \cite{Ecker:2002cw}
 predictions (solid and dotted lines) in low energy region
 and experimental data \cite{Akhmetshin:1998df,Achasov:2003bv,bb1}.
}
\label{neutralEcker}
\end{figure}
\begin{figure*}[ht] 
\includegraphics[width=7.6cm,height=4.3cm]{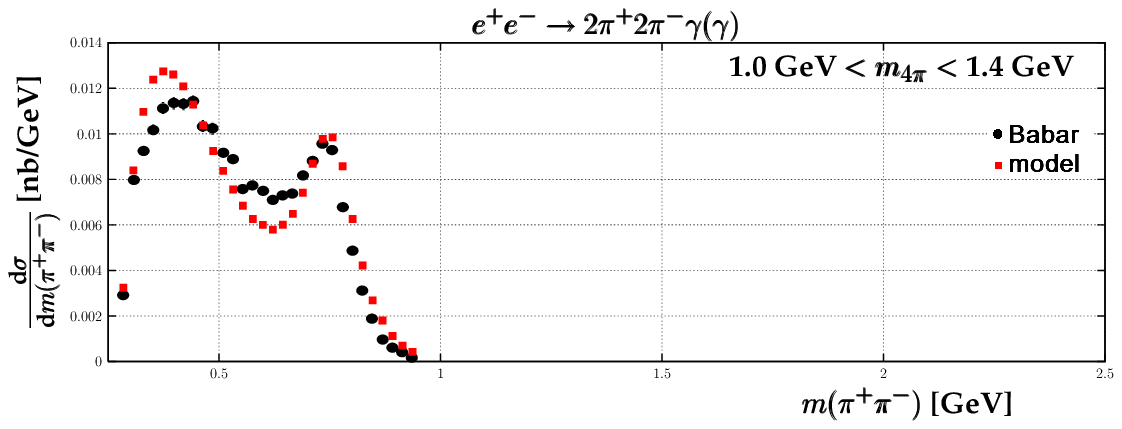}\kern +30pt
\includegraphics[width=7.6cm,height=4.3cm]{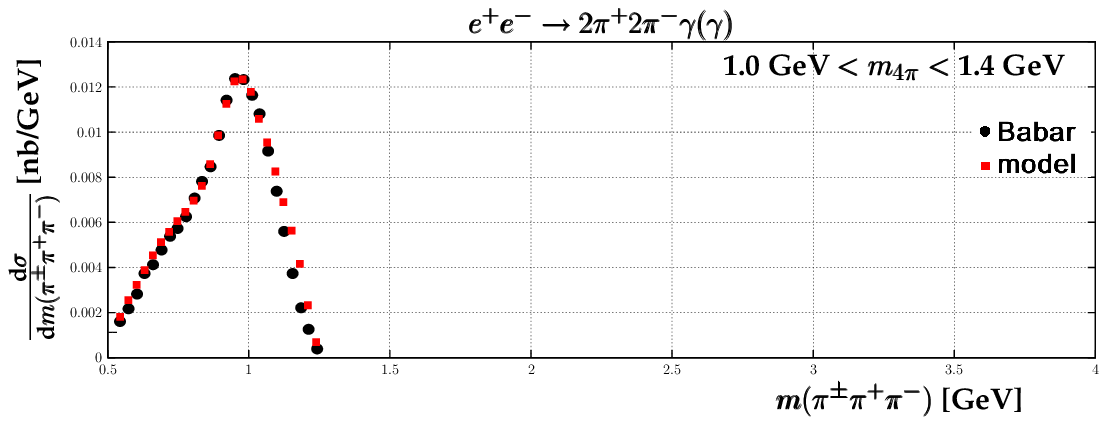}\\
\includegraphics[width=7.6cm,height=4.3cm]{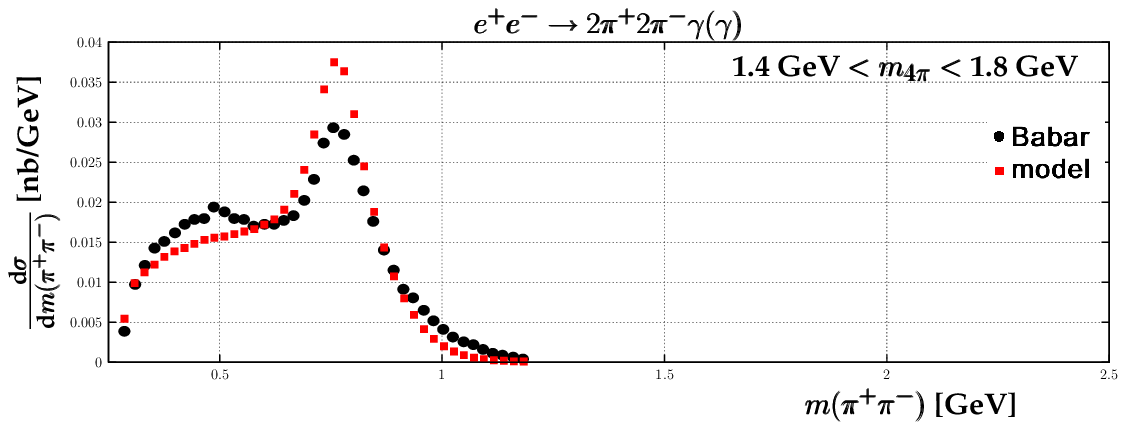}\kern +30pt
\includegraphics[width=7.6cm,height=4.3cm]{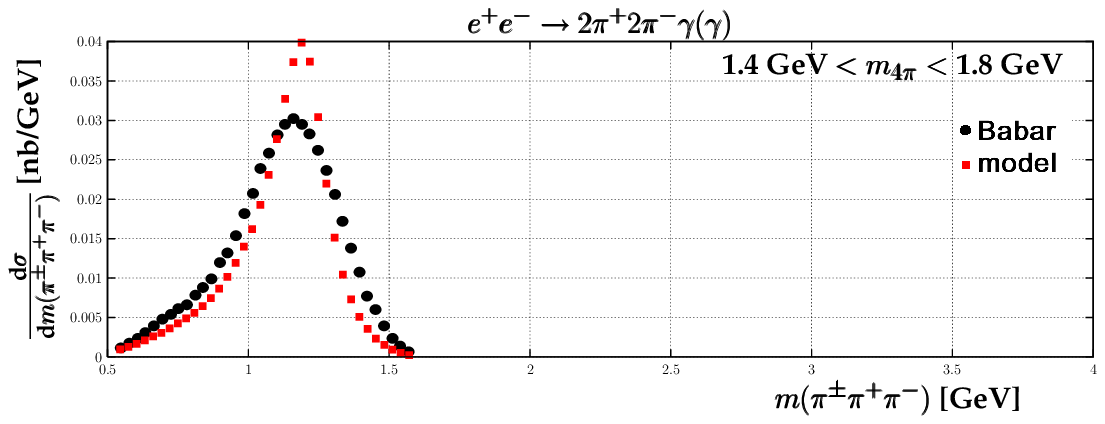}\\
\includegraphics[width=7.6cm,height=4.3cm]{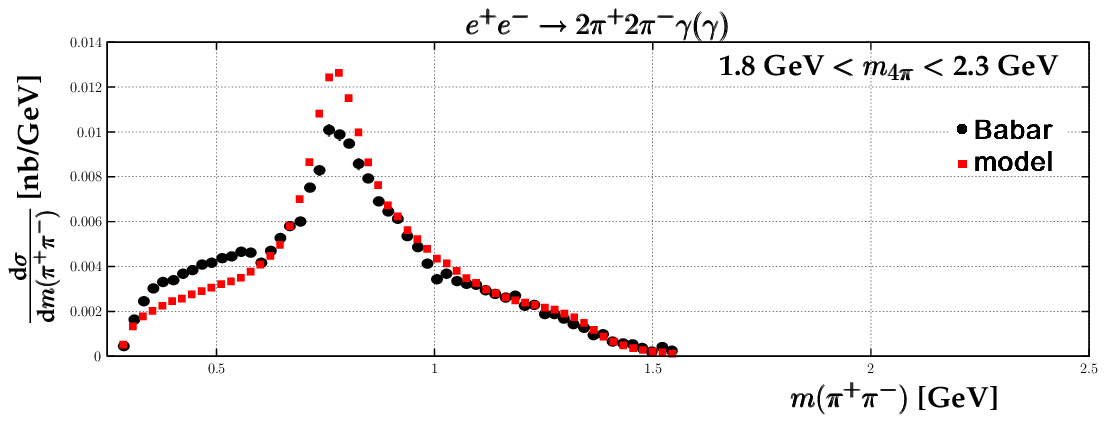}\kern +30pt
\includegraphics[width=7.6cm,height=4.3cm]{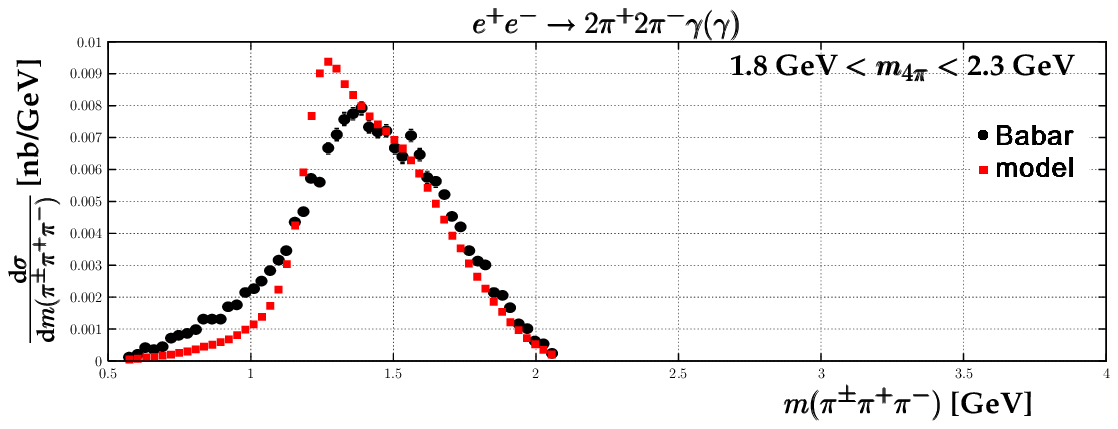}\\
\includegraphics[width=7.6cm,height=4.3cm]{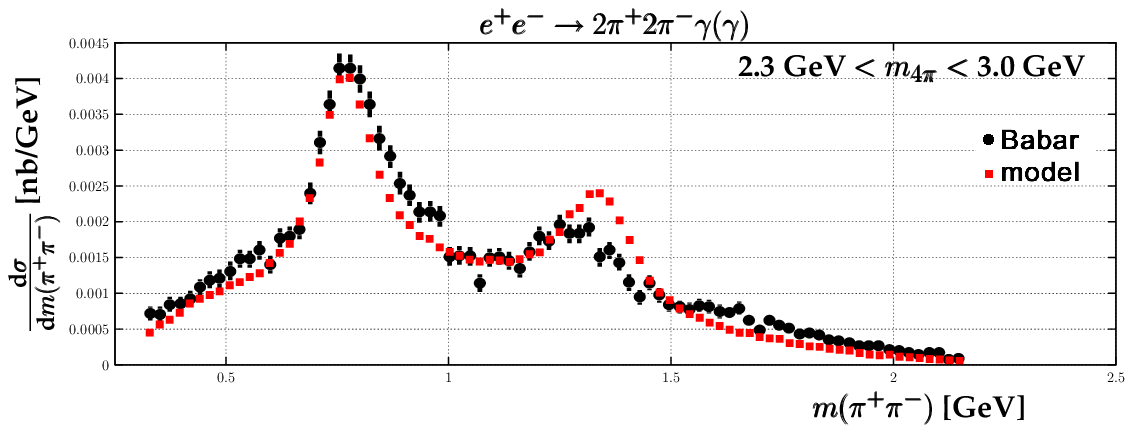}\kern +30pt
\includegraphics[width=7.6cm,height=4.3cm]{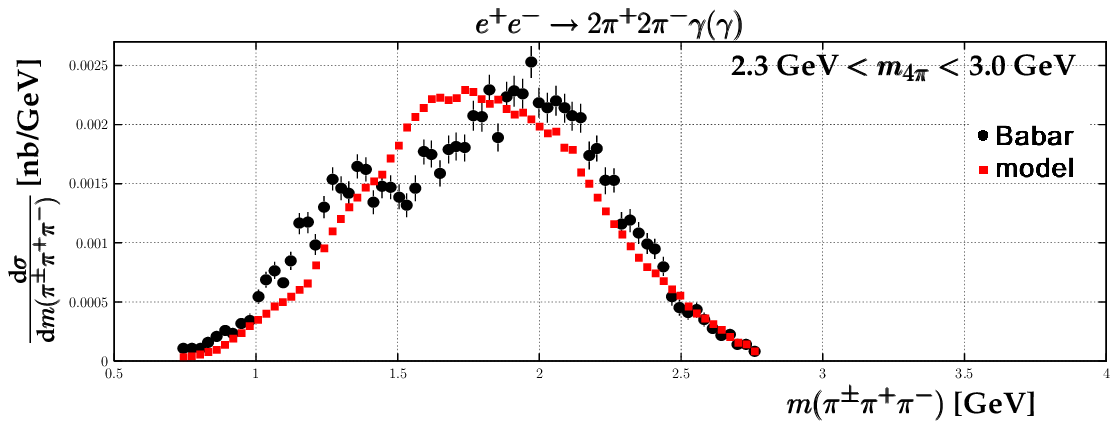}\\
\includegraphics[width=7.6cm,height=4.3cm]{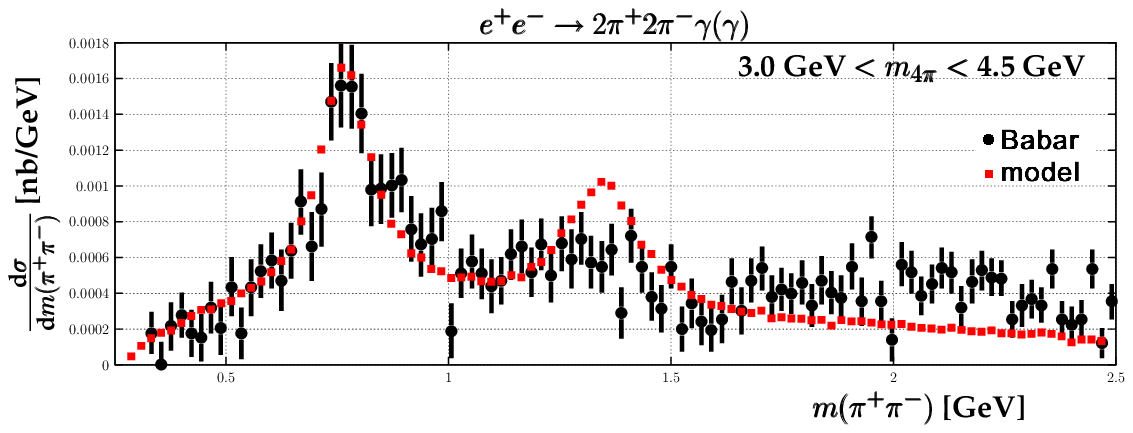}\kern +30pt
\includegraphics[width=7.6cm,height=4.3cm]{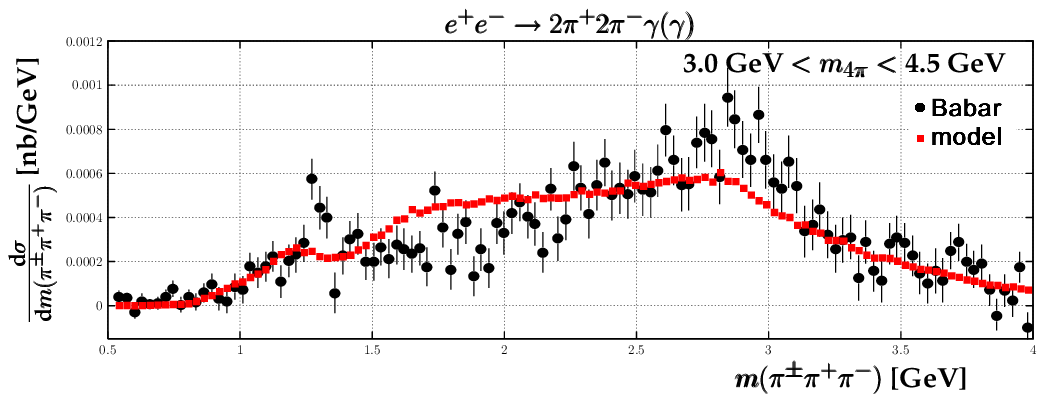}
\caption{(color online). 
Two and three pion invariant mass distributions for five different
 ranges of   $2\pi^+2\pi^-$ invariant mass. 
  The BaBar data points (filled circles), given as events/bin,
 are superimposed
 on plots obtained by PHOKHARA (filled squares)
 (see text for details).
 }
 \label{subdistr}
\end{figure*}
\begin{figure*}[ht] 
\begin{center}
\includegraphics[width=7.6cm,height=4.3cm]{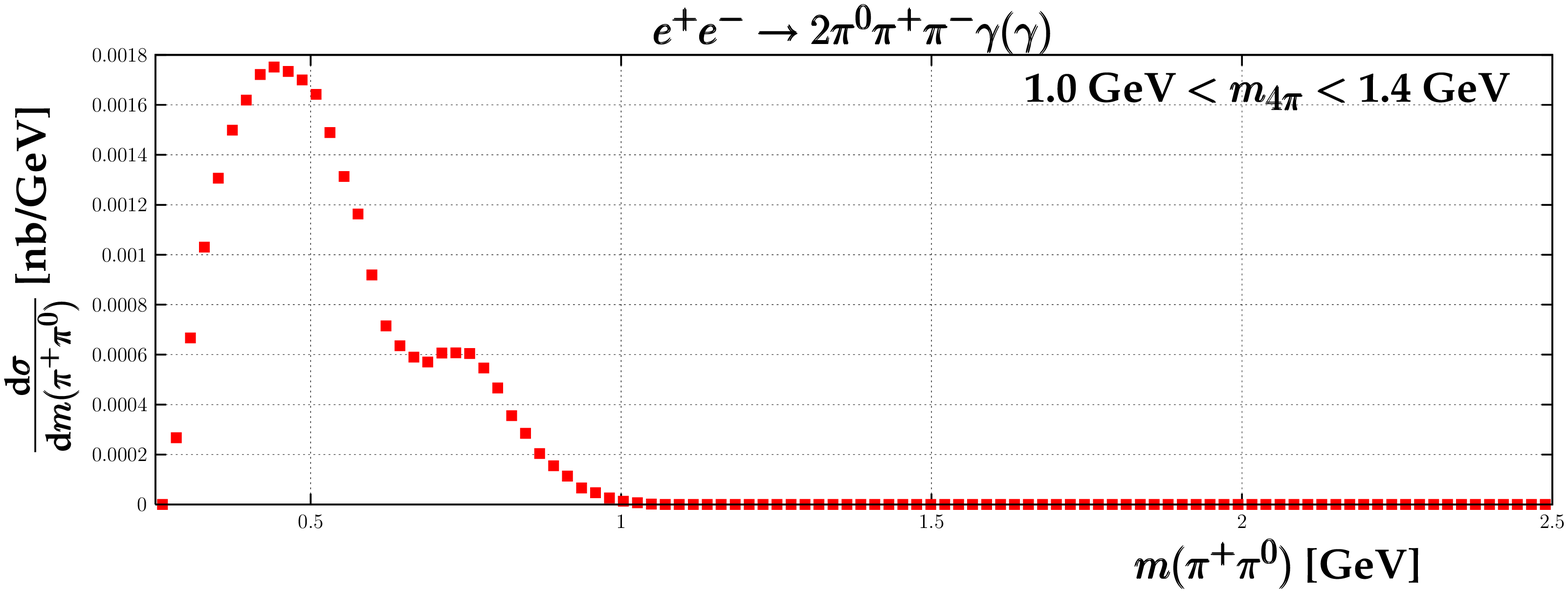}
\includegraphics[width=7.6cm,height=4.3cm]{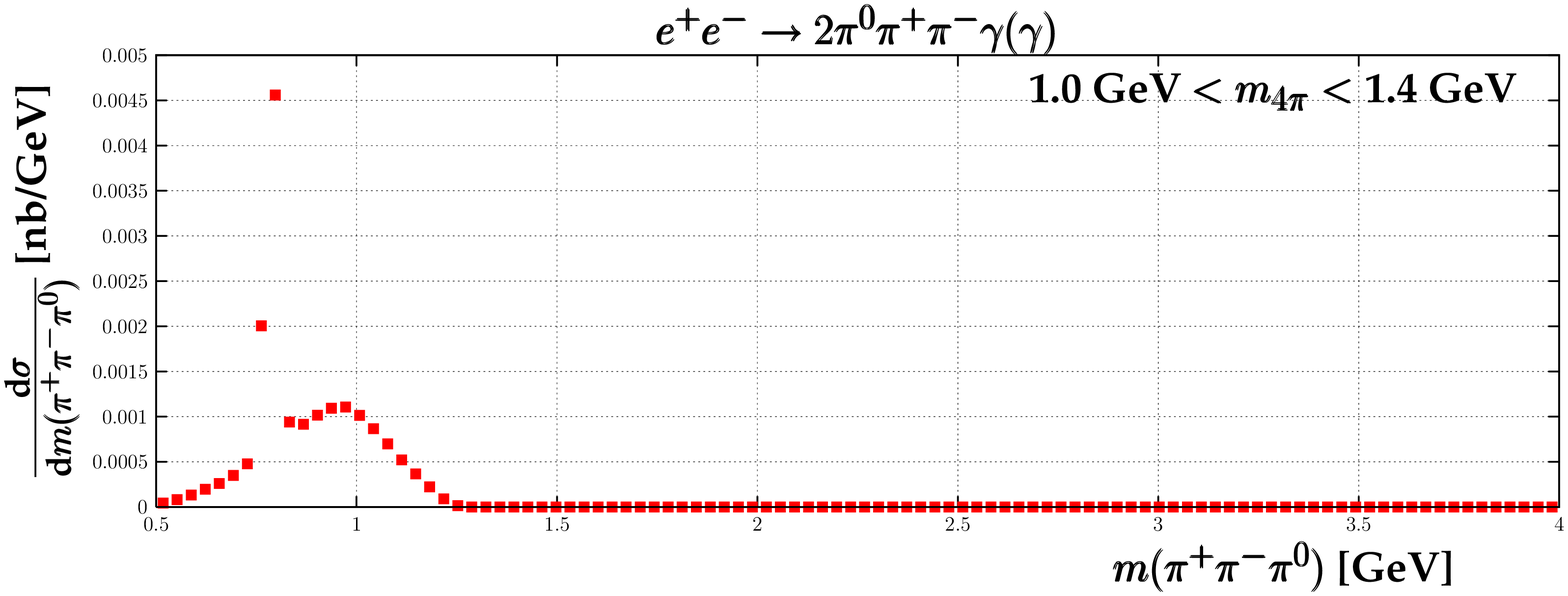}\\
\includegraphics[width=7.6cm,height=4.3cm]{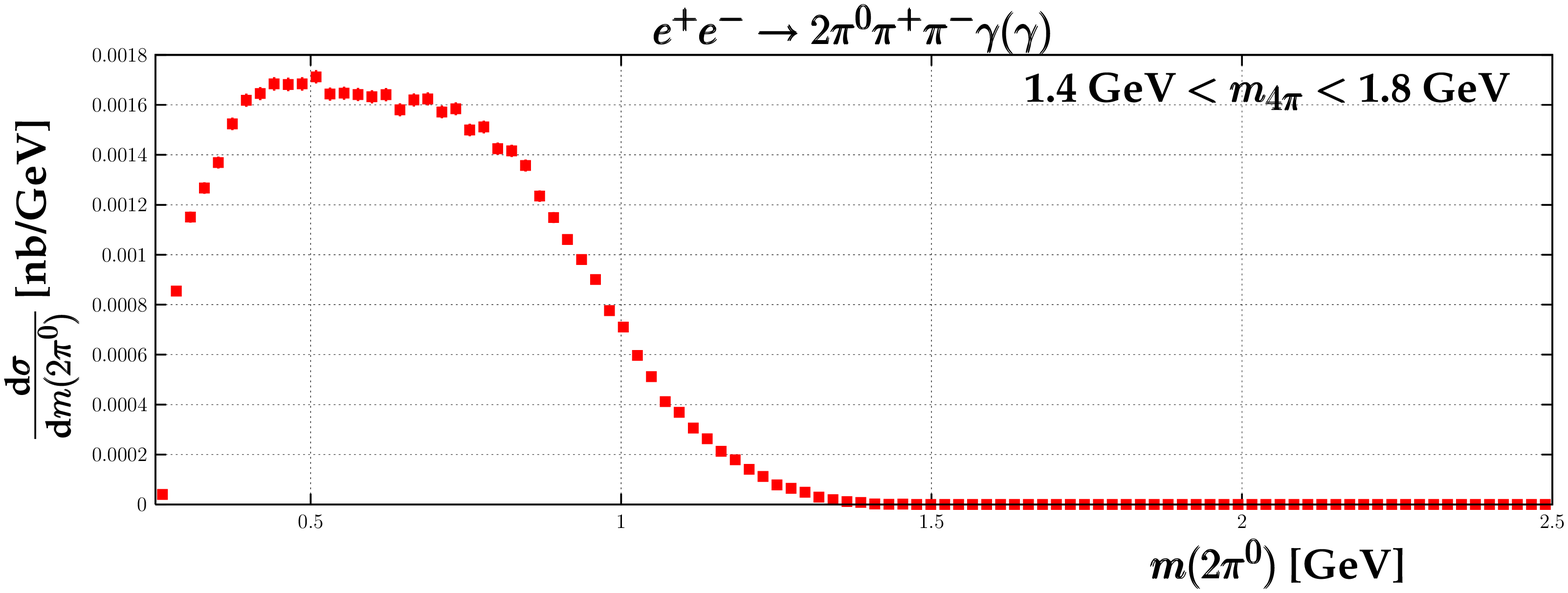}
\includegraphics[width=7.6cm,height=4.3cm]{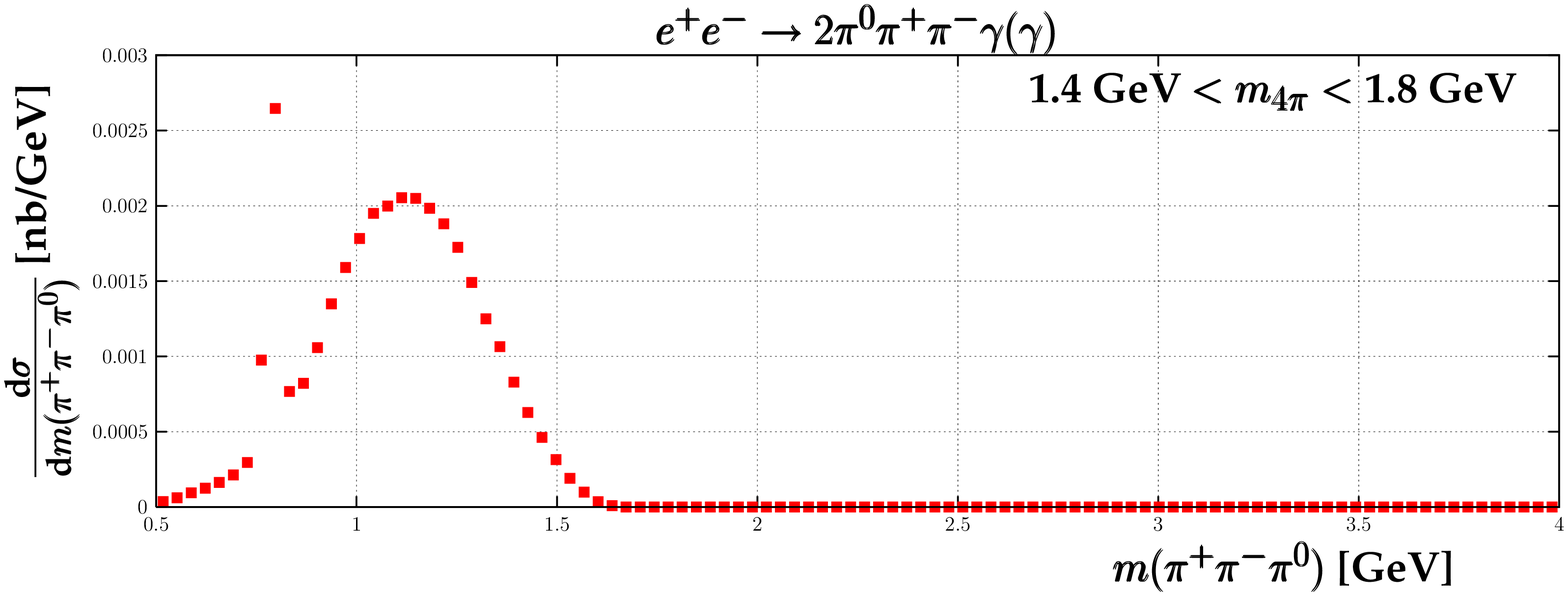}\\
\includegraphics[width=7.6cm,height=4.3cm]{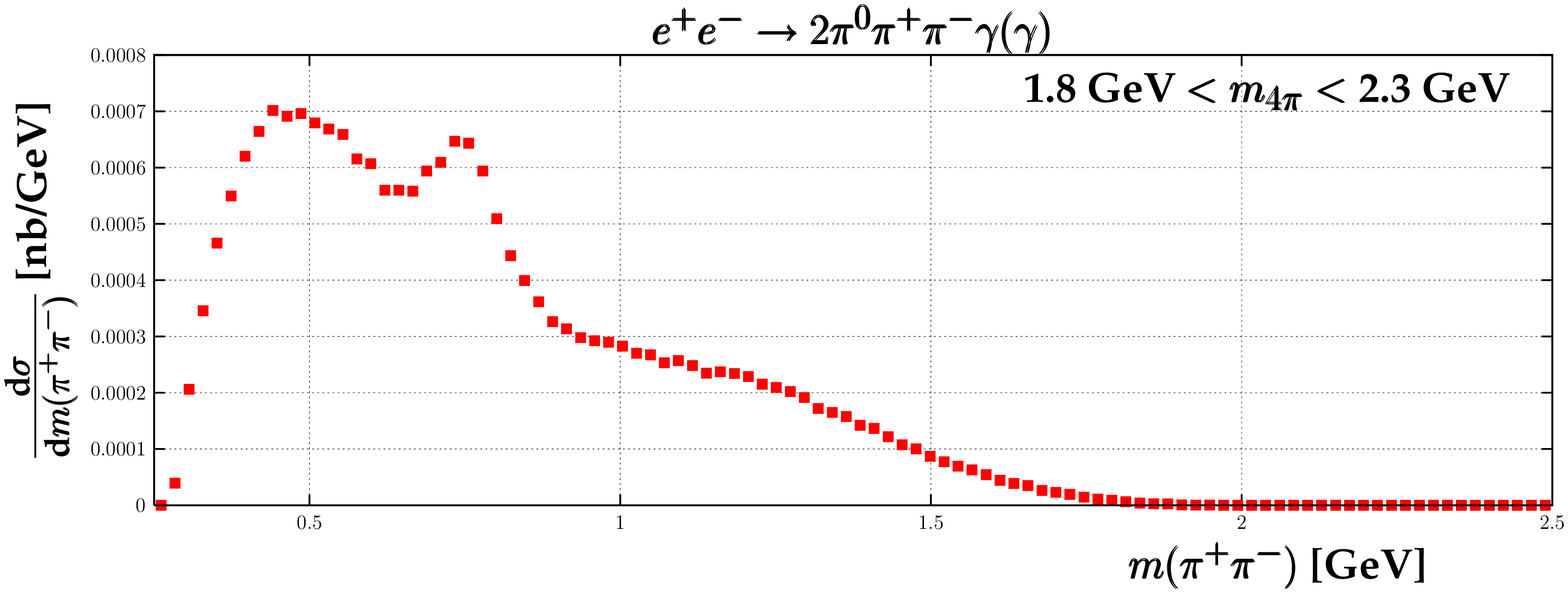}
\includegraphics[width=7.6cm,height=4.3cm]{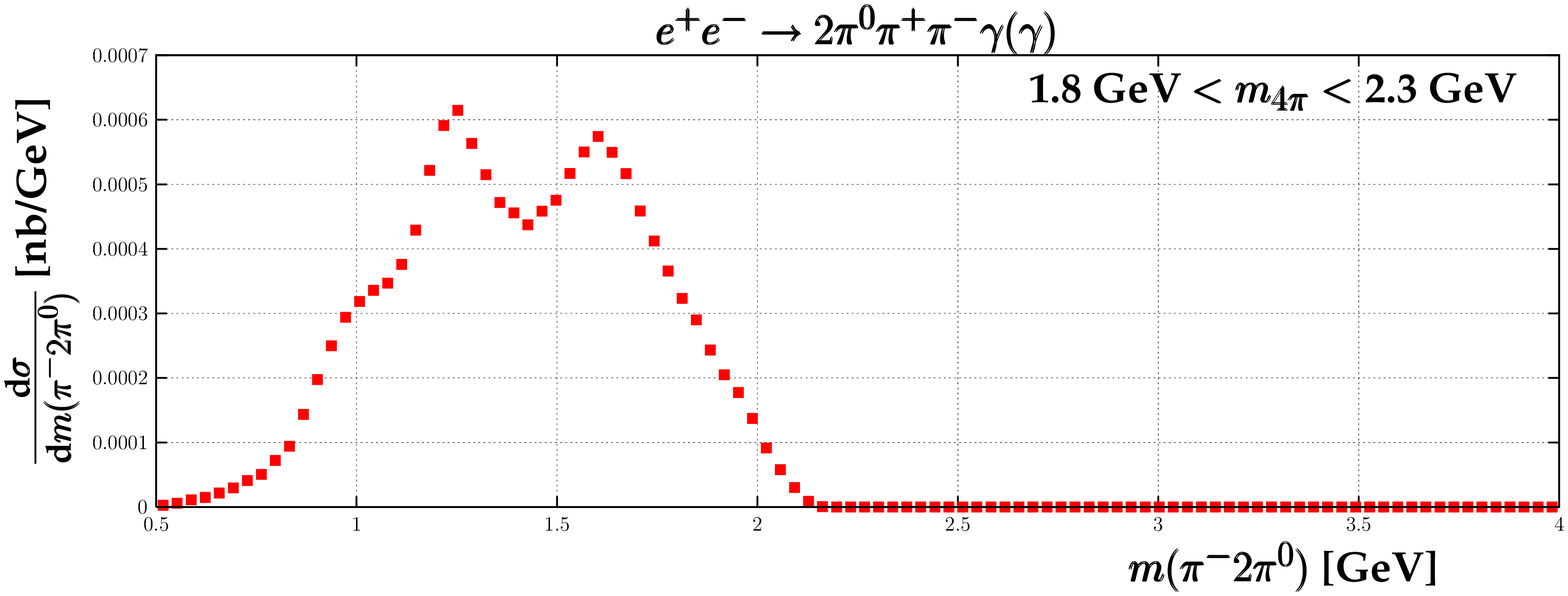}\\
\includegraphics[width=7.6cm,height=4.3cm]{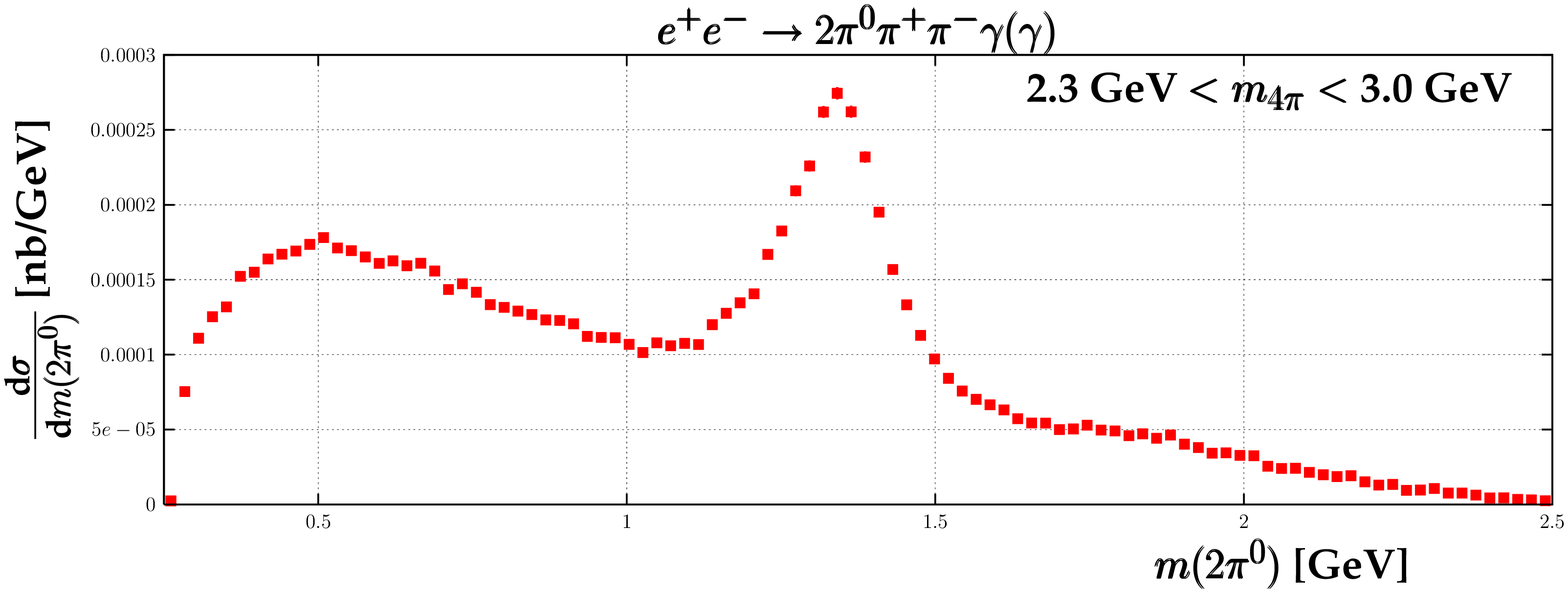}
\includegraphics[width=7.6cm,height=4.3cm]{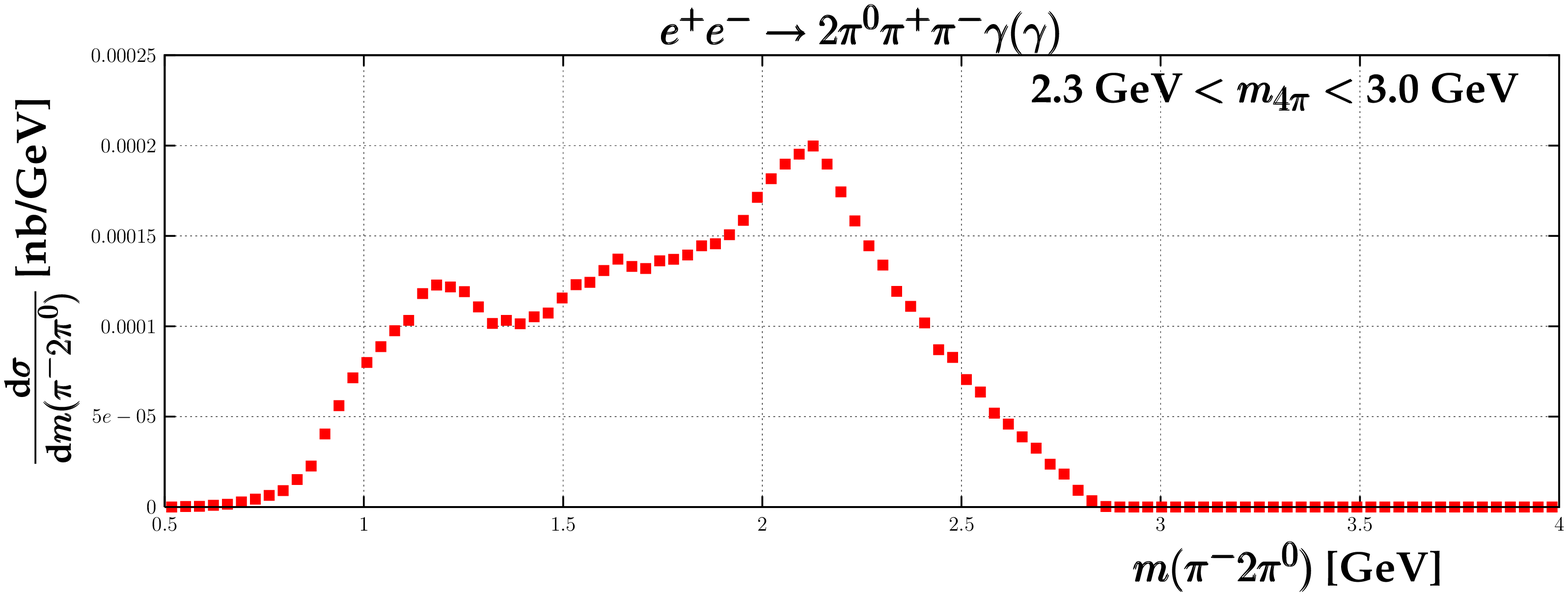}\\
\includegraphics[width=7.6cm,height=4.3cm]{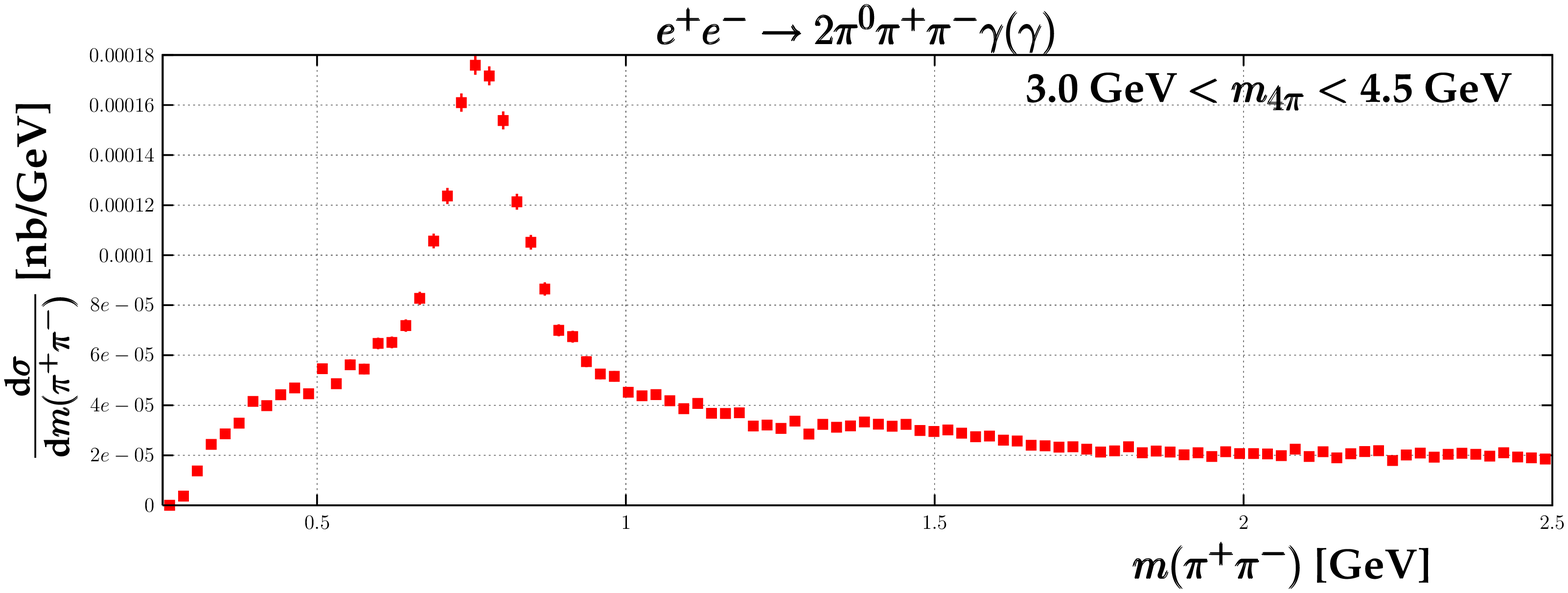}
\includegraphics[width=7.6cm,height=4.3cm]{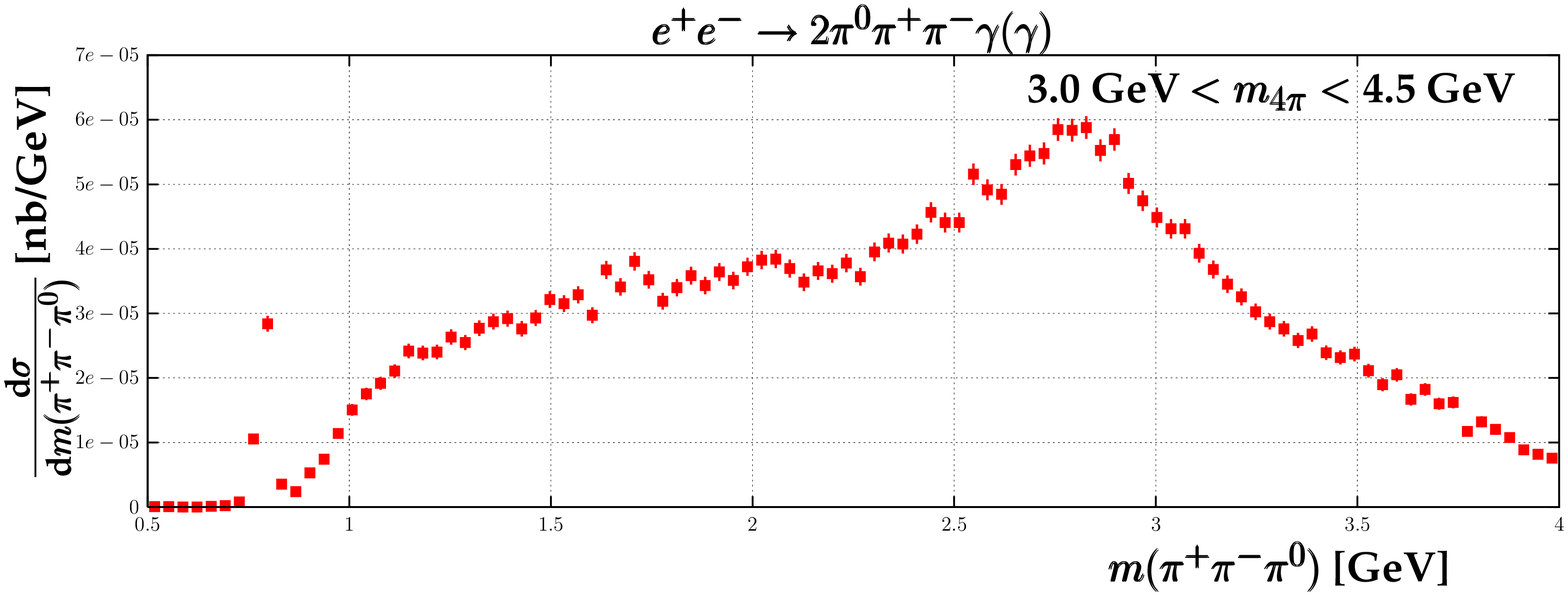}
\caption{(color online). 
Predicted by the model
 two and three pion invariant mass distributions for five different
 ranges of   $2\pi^0\pi^+\pi^-$ invariant mass (selected set). 
}
 \label{subdistr_neutr}
\end{center}
\end{figure*}

   The results are shown in Figs. \ref{omegafit},
 \ref{neutralfit} and \ref{chargedfit}
    and in Table \ref{babarfit}. The fit is quite good,
  with $\chi^2 / n_{d.o.f} = 275 / 287$. However, one has to remember 
  that only the cross sections were fitted and all sub-distributions
  are to large extent determined by the model assumptions. 
 The constants $\beta^{a_1}_i$, $\beta^{f_0}_i$ and  $\beta^{\omega}_i$
 (with $i=1,2,3$) characterize the relative importance of the radial
  $\rho$ excitations (compared to the one of the ground state,
  $\beta_0\equiv 1$)
  in the amplitudes depicted in Figs. \ref{origc}, and \ref{omega}
 (see also \Eq{frhodef}).
 Large values of 
  $\beta^{f_0}_i$  are the consequence of the small 
  $f_0(1370)-\rho_0-\rho_0$ coupling compared to higher $\rho$ radial
  excitations, which are indeed dominated by $\rho_1$.

 It is interesting to see how the model compares 
 to predictions based on the chiral Lagrangian \cite{Ecker:2002cw}
 in the low $Q^2$ region, where this ansatz is expected to be applicable.
 In Fig. \ref{chargedEcker} (Fig. \ref{neutralEcker})
  this comparison is shown for the charged (neutral) mode, together
 with data from BaBar \cite{Aubert:2005eg,bb1},
 CMD2 \cite{Akhmetshin:1998df,Akhmetshin:1999ty,Akhmetshin:2004dy}
 and SND \cite{Achasov:2003bv}.
 Since our model parameters were fitted to that cross section 
 the thick dotted curve is not a prediction,
 apart of the $\sqrt{Q^2}$ region
 below 0.8 GeV, from where the contribution to $\chi^2$ of the fit is
 negligible due to
 the low accuracy of the data.

The sub-distributions can be qualitatively
 compared (Fig. \ref{subdistr}) with plots presented by BaBar
\cite{Aubert:2005eg}. These were not used in the fit and thus can 
 be considered as predictions. 
 Integrals for both experimental and theoretical plots
 are equal by construction.
  Further refinements
 of the model will be possible when the data on sub-distributions will
 become available.

The contributions from two $\rho$ mesons in the final state are shown as
dashed line
  in Fig. \ref{neutralfit}. They were extracted selecting events
  with $\pi^+\pi^0$ and $\pi^-\pi^0$ invariant masses within 
  the range from $m_{\rho}-\Gamma_{\rho}$ to $m_{\rho}+\Gamma_{\rho}$.
  These are affected by background from the other amplitudes and thus
  do not correspond exactly to the contributions from 
  $\rho$ part of the current (Fig. \ref{rhorho} and \Eq{rhorhocurrent}),
  hence the separation is not as clean as for the 
  $\omega$ case. The model prediction is
  smaller than the BaBar result
 \cite{bb1}.

 Selected two and three pion invariant mass sub-distributions for
 the reaction $e^+e^- \to 2\pi^0\pi^+\pi^-\gamma(\gamma)$ are shown in 
 Fig. \ref{subdistr_neutr}. The contributions from various resonances 
 included in the model are clearly visible. Comparisons of the predictions
 will be possible, when the final BaBar results are published.

\section{Model predictions for $\tau$ decays }\label{taudecays}

 One can confront the model with the  data \cite{Yao:2006px} for
  the partial $\tau$ decay rates to four pion 
 final states. The results  are collected in Table \ref{Branchingratios}.
 The theoretical error is obtained from the errors of the model parameters
 extracted in the fit. 
 Within the quoted errors, the predictions are in good agreement
 with the data 
 even if 
 one observes sizable difference between the data for
  $Br(\tau^-\to\nu_\tau2\pi^-\pi^+\pi^0)$ and the prediction
 via the isospin relations.
 At present the results are still consistent within the conservatively
 estimated error, which is dominated by the one of the preliminary
 BaBar result for $\sigma(e^+e^-\to 2\pi^0\pi^+\pi^-)$.
 With an expected error of about 5\%, the final BaBar result
 will further push the accuracy of the isospin symmetry tests.
 
The model of $4\pi$ hadronic current proposed in this paper
  was fitted to BaBar data and  relies on isospin
 symmetry. Thus its predictions for the $\tau$ spectral functions
 follows the predictions from BaBar data based on the isospin symmetry
 assumption (presented in Section \ref{expsituation}),
 apart from few percent phase space effects coming
 from the $\pi^\pm-\pi^0$ mass difference. The model predictions are
 also shown in 
 Fig. \ref{spec1}, \ref{spec2} and \ref{spec3}.
 The central curve represents the model predictions, the upper 
 and lower curves are the error estimates based on errors coming from the
  fitted parameters of the model.
 
\begin{figure}[ht] 
\begin{center}
\includegraphics[width=3.8cm,height=2.15cm]{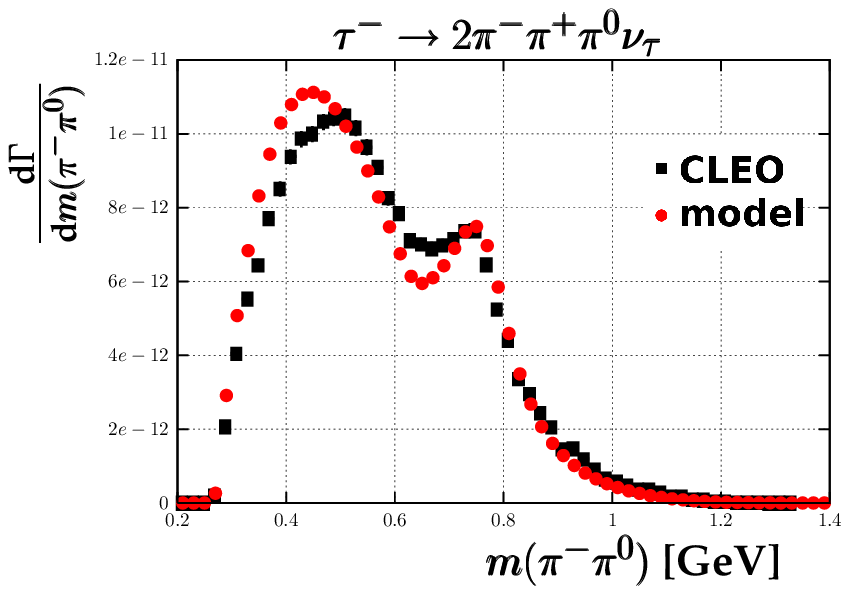}
\includegraphics[width=3.8cm,height=2.15cm]{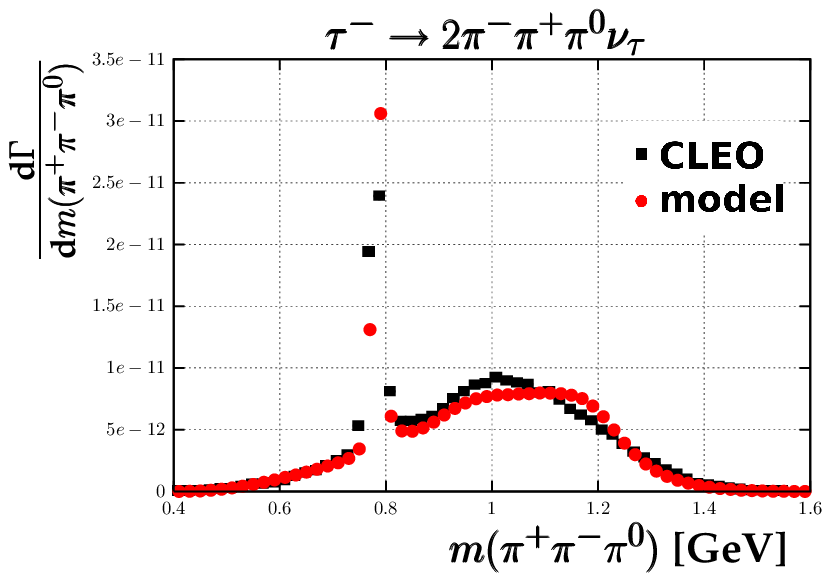}\\
\includegraphics[width=3.8cm,height=2.15cm]{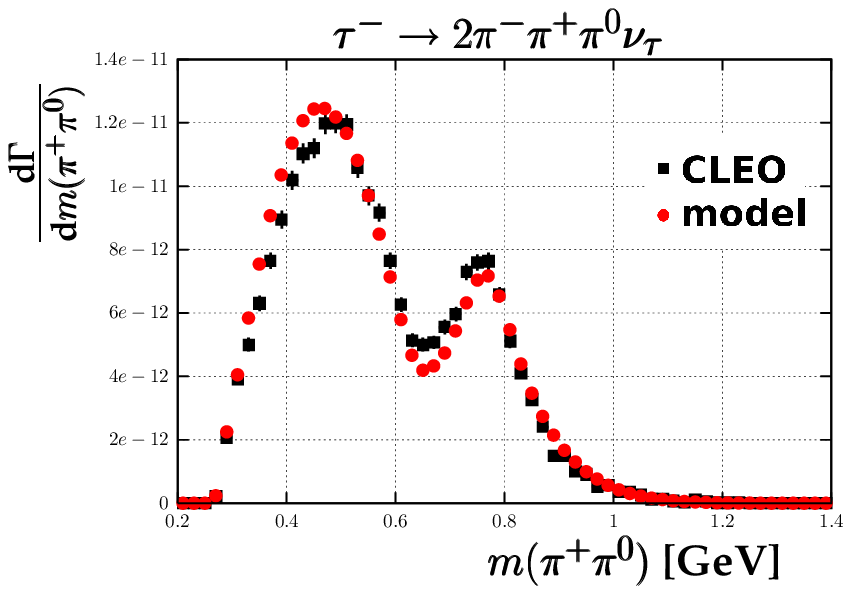}
\includegraphics[width=3.8cm,height=2.15cm]{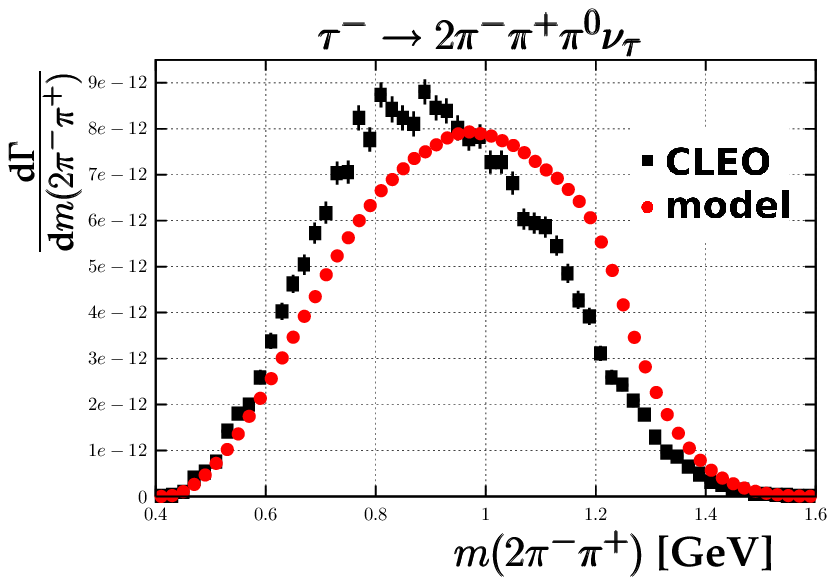}\\
\includegraphics[width=3.8cm,height=2.15cm]{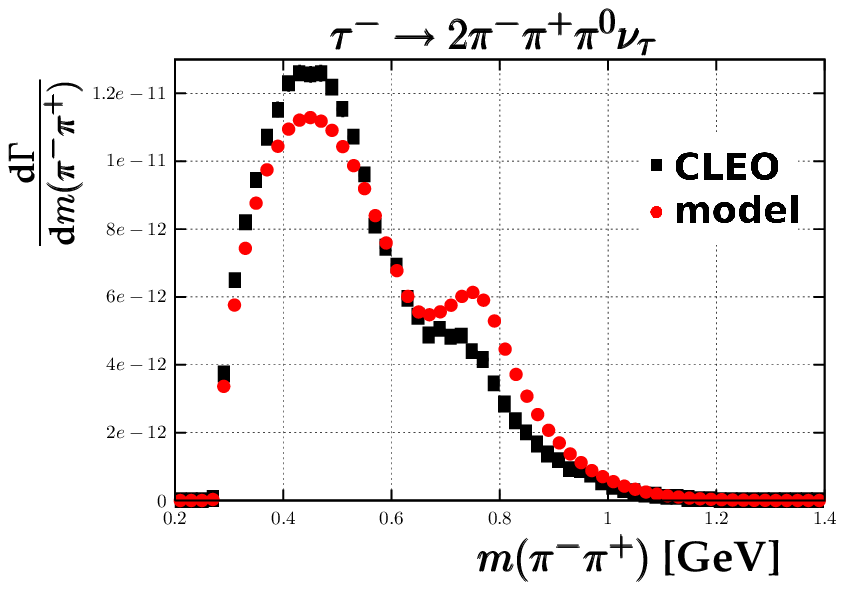}
\includegraphics[width=3.8cm,height=2.15cm]{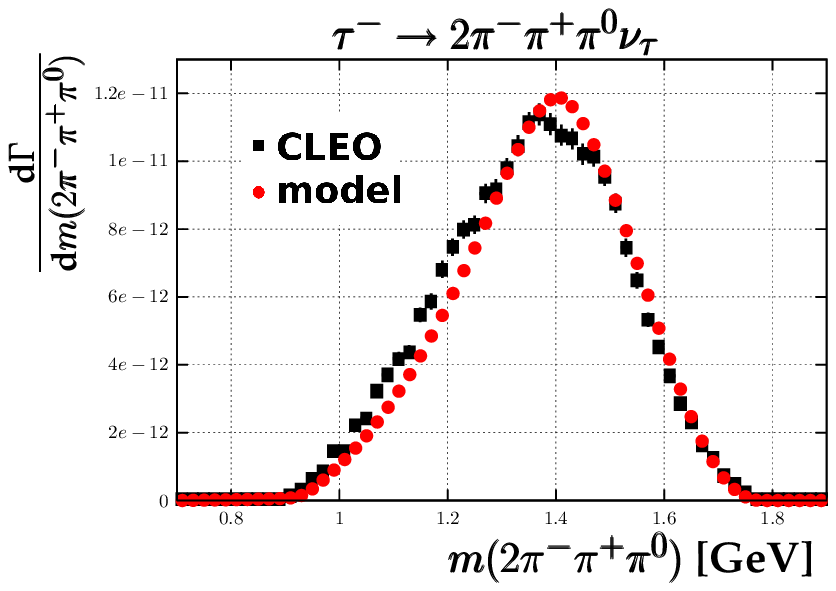}\\
\caption{(color online). 
Two and three pion invariant mass sub-distributions for different
 pion charge combinations. 
  The CLEO data points (filled circles), given as events/bin,
 are superimposed
 on plots obtained within studied model (filled squares).
 See text for details. }
 \label{subdistrCLEO}
\end{center}
\end{figure}

Two, three and four pion invariant mass distributions obtained within
 our model are compared with CLEO data (available only as plots)
 in Fig. \ref{subdistrCLEO}. Although predictions and the data differ as far 
 as the detailed description is concerned, good qualitative agreement
  is observed.

\begin{widetext}

\begin{table}[ht]
\begin{tabular}{c|c|c|c}
\hline
  & Br($\tau-\to\nu_\tau2\pi^-\pi^+\pi^0$) & Br($\tau-\to\nu_\tau\pi^-\omega(\pi^-\pi^+\pi^0)$) & Br($\tau-\to\nu_\tau\pi^-3\pi^0$)\\
\hline
 PDG \cite{Yao:2006px} &(4.46$\pm$ 0.06)\% &(1.77$\pm$ 0.1)\% & (1.04$\pm$ 0.08)\% \\
\hline
 model &(4.12 $\pm$ 0.21)\% & (1.60 $\pm$ 0.13)\% & (1.06 $\pm$ 0.09)\% \\
\hline
 BaBar (CVC) & (3.98 $\pm$ 0.30)\%& (1.57 $\pm$ 0.31)\% & (1.02$\pm$ 0.05)\% \\
\hline
\end{tabular}
\caption{Branching ratios of $\tau$ decay modes. Results 
 of our model are compared to experimental data \cite{Yao:2006px}
 and predictions based on BaBar data \cite{Aubert:2005eg,bb1}
 and isospin symmetry}
\label{Branchingratios}
\end{table}
\end{widetext}

\section{Implementation into PHOKHARA and tests of the Monte Carlo generator}
\label{PHOKHARA}

 The model for the  hadronic current was implemented into the PHOKHARA 
 event generator (version 7.0). It will be available at
 {\tt http://ific.uv.es/~rodrigo/phokhara/} together with the implementation
  of the $J/\psi$ and $\psi(2S)$ contributions to 2-body hadronic
 final states (in preparation).
 Only the current $J_{\mu}$ was coded 
 in the form described in the Appendix;  the charged mode
 is obtained via the relation Eq. (\ref{rel}). Neither the
 $\omega$- part of the current nor the double $\rho$ resonance diagrams
 (left in Fig. \ref{rhorho}) contribute to that part. Only a priori weights,
  used in the multi-channel Monte Carlo generation, were changed
  as compared to previous versions \cite{Czyz:2002np}.
 Nevertheless,
 tests checking the implementation were performed to assure a proper technical
 precision of the code. The NLO version of the code was checked for the
 configurations without any cuts against
 analytic NLO results of \cite{Berends:1987ab} (see also \cite{Czyz:2002np}),
 separately for one-- and two-- photon contributions. The separation
 $w=\frac{E_\gamma}{\sqrt{s}}=10^{-4}$ between soft (integrated analytically)
 and hard (generated) parts was used in this test.
 The precision of the tests, limited by the  Monte Carlo statistics,
 is significantly below one per mill. 
 As the analytic formula
 contains as a factor the cross section of the process without photon
 emission ($\sigma(e^+e^-\to 4\pi)$), which is not known analytically,
 it was obtained be means of Monte Carlo integration by a dedicated program.
 In that program, in contrast to PHOKHARA,
  flat phase space generation was used to avoid any errors
 due to the change of variables. 
 The independence of the results of the generation on the separation
 into soft and hard parts was also tested with similar precision.

\section{Conclusions} \label{conclusions}

Four-pion production in $e^+e^-$ annihilation and in $\tau$ decays is
characterized by a multitude of resonant sub-channels. This makes it
difficult to construct an amplitude which is based on first principles of
QCD only. In this paper we have constructed a model amplitude which
incorporates a limited set of channels, namely $a_1\pi$, $\rho f_0$,
$\rho\rho$ and $\omega\pi$ and which is approximately consistent with
chiral predictions for small $Q^2$. A number of parameters which
characterize the relative importance of the various couplings and of the
radial excitations of the $\rho$ meson is fitted to the cross sections
for $2\pi^+2\pi^-$ and $\pi^+\pi^-2\pi^0$ as measured by the BaBar
collaboration. The model predictions for the two- and three-pion mass
distributions which are not fitted separately, are consistent with the
data both from $e^+e^-$ annihilation and from $\tau$ decays. Furthermore,
we find that the present data are, within their 5~--~10\% systematic error,
consistent with the relations derived from isospin invariance, and which
are intrinsic for our model. The amplitude is incorporated into the
Monte Carlo generator PHOKHARA, simulating $4\pi$ production through the
radiative return.

\begin{acknowledgments}
We are grateful to Achim Denig for numerous discussions 
 and reading of the manuscript.
Henryk Czy\.z and Agnieszka Wapienik are
  grateful for the support and the kind hospitality
of the Institut f{\"u}r Theoretische Teilchenphysik
 of the Karlsruhe University.

\end{acknowledgments}

\appendix*

\section{The current}

In this appendix we give a complete definition of the hadronic current used
in this paper. We take as a basic building block the electromagnetic
 current $J_\mu^{em}= \frac{1}{\sqrt{2}} J^3_{\mu}$ (using $J^{I=0}_\mu =0$),
 which is of direct
 importance for implementation in PHOKHARA and define
 $ \Gamma^{\mu}\equiv \frac{1}{\sqrt{2}}\langle \pi^+ \pi^- \pi_1^0 \pi_2^0 | J^3_{\mu} | 0 \rangle$.
 Other channels can be obtained through \Eq{rel}.
 
 $\Gamma^{\mu}$ can be decomposed into the
 following four parts:
\beq
\Gamma^{\mu} = \Gamma^{\mu}_{a_1} + \Gamma^{\mu}_{f_0}+
\Gamma^{\mu}_{\omega}+ \Gamma^{\mu}_{\rho}.
\label{currentsplit}
\eeq
We denote the four pion momenta 
by $q{_1}(\pi^0),\ q{_2}(\pi^0),\  q{_3}(\pi^-)$ 
and $q{_4}(\pi^+) $
and use the proper pion masses $m_{\pi^{\pm}} $ and $m_{\pi^0}$
 wherever appropriate.
Thus the current possesses  isospin
symmetry,  broken only by kinematic effects.
The general structure is largely based on \cite{Czyz:2000wh}.
Contribution from the part containing an $a_1$
exchange reads:
\begin{eqnarray}
&&\kern-20pt\Gamma^{\mu}_{a_1}\left(q_1,q_2,q_3,q_4\right) = 
\nonumber\\\phantom{}\kern-60pt &&\tilde \Gamma^{\mu}_{a_1}(q_3,q_2,q_1,q_4)
 +\tilde \Gamma^{\mu}_{a_1}(q_3,q_1,q_2,q_4) \nonumber\\
\phantom{}\kern-60pt&-& \tilde \Gamma^{\mu}_{a_1}(q_4,q_2,q_1,q_3)
- \tilde \Gamma^{\mu}_{a_1}(q_4,q_1,q_2,q_3). 
 \end{eqnarray}
 
\noindent
The function  $\tilde \Gamma^{\mu}_{a_1}$ is of the form
\begin{eqnarray}
 &&{\kern -15pt}
\tilde \Gamma^{\mu}_{a_1}(q_1,q_2,q_3,q_4) =
  c_{a_1}\ F_{\rho}\left(Q^2,\vec \beta^{a_1}\right)  \ 
  BW_{a_1}\left((Q-q_1)^2\right) \ \nonumber \\
&&\times B_{\rho}\left((q_3+q_4)^2\right) \ \Bigg[(q_3-q_4)^{\mu}+ 
 q_1^{\mu} \frac{q_2(q_3-q_4)}{(Q-q_1)^2} \nonumber \\
&& -Q^{\mu}\left(\frac{Q(q_3-q_4)}{Q^2}
 +\frac{(Q q_1)(q_2(q_3-q_4))}{Q^2 (Q-q_1)^2}\right)\Bigg],\nonumber \\
\end{eqnarray}
and corresponds to the configuration $\rho(Q)\to \pi(q_1)\  a_1(\to\rho\ \pi(q_2))$
with $\rho\to\pi(q_3)\ \pi(q_4)$. The other three terms are enforced
 by Bose symmetry ($q_1\to q_2$) and charge conjugation.

 The contribution from 
$\rho \to f_0(\pi(q_1)\ \pi(q_2)) \rho(\to \pi(q_3)\ \pi(q_4))$ reads

\begin{eqnarray}
 &&{\kern -25pt}\Gamma^{\mu}_{f_0}(q_1,q_2,q_3,q_4)  =
   c_{f_0} \ F_{\rho}\left(Q^2,\vec \beta^{f_0}\right) \ T_{\rho}\left((q_3+q_4)^2\right) \ \nonumber \\
 &&{\kern -20pt}BW_{f_0}\left((q_1+q_2)^2\right)
\left[(q_3-q_4)^{\mu} - Q^{\mu} \frac{Q(q_3-q_4)}{Q^2}\right] \  .
\end{eqnarray}


 The contribution coming from  the anomalous  part of the current
 (containing \(\omega\)
 exchange) reads

\beq
\Gamma^{\mu}_{\omega}(q_1,q_2,q_3,q_4) = \tilde{\Gamma}^{\mu}_{\omega}(q_1,q_2,q_3,q_4) 
 +  \tilde{\Gamma}^{\mu}_{\omega}(q_2,q_1,q_3,q_4)
\label{omega_c}
\eeq
\noindent
with
\bea
\kern-2pt\tilde{\Gamma}^{\mu}_{\omega}(q_1,q_2,q_3,q_4)&=&
 2\ c_{\omega} \ g_{\omega\pi\rho}\ g_{\rho\pi\pi}\ 
 F_{\rho}(Q^2,\vec \beta^{\omega}) \nonumber \\
&{\kern -160pt}\times& {\kern -85pt}BW_{\omega}((Q-q_1)^2)\ 
   H_{\rho}((q_2+q_3)^2,(q_2+q_4)^2,(q_3+q_4)^2)\nonumber \\
&{\kern -16pt}\times& {\kern -10pt}[q_2^{\mu}((q_1q_4)(q_3Q)-(q_1q_3)(q_4Q))\nonumber \\
         &{\kern -16pt}+& {\kern -10pt} q_3^{\mu}((q_1q_2)(q_4Q)-(q_1q_4)(q_2Q))\nonumber\\
	   &{\kern -16pt}+&{\kern -10pt} q_4^{\mu}((q_1q_3)(q_2Q)-(q_1q_2)(q_3Q))].
\eea
\noindent
where
\beq
g_{\omega\pi\rho}= 42.3\ {\rm GeV}^{-5},\ \ \ g_{\rho\pi\pi}=5.997.
\eeq
The first term in \Eq{omega_c}  corresponds to the configuration 
$\rho\to \pi(q_1)\omega(\to \pi(q_2)\pi(q_3)\pi(q_4))$
 the second follows from the Bose symmetry.

The structure of the omega decay was taken from \cite{Czyz:2005as}.
Other possible contributions to the $3\pi$ part of the current
coming from $\phi(1020)$ or higher radial
$\omega$ excitations are not seen in the data and thus were not included
into the model.

 $\Gamma^{\mu}_{\rho}$ part of the current is of the
 following form:


\bea
&&{\kern -20pt}\Gamma^{\mu}_{\rho}(q_1,q_2,q_3,q_4) =  \nonumber\\
 &&\kern -10pt c_{\rho} \ g_{\rho\pi\pi}^3 \ g_{\rho\gamma}
 \ BW^{\rho_0,\rho_1}(Q^2) \times 
 \left(g^\mu_\nu -\frac{Q^\mu Q_\nu}{Q^2}\right)\times \nonumber\\
&&\Big\{ \Big[  \left(G_\rho^\mu(q_1,q_2,q_3,q_4)+G_\rho^\mu(q_4,q_1,q_2,q_3)
 \right)
-\left(3\leftrightarrow 4\right) 
 \Big]\nonumber\\
&&+\Big[1\leftrightarrow 2\Big]\Big\}
\label{rhorhocurrent}
\eea
\noindent
where
\bea
&&\kern-26ptG_\rho^\mu(q_1,q_2,q_3,q_4)= 
 q_1^\mu BW^{\rho_0,\rho_1}((q_1+q_3)^2)\nonumber \\
&&\kern-16pt\times\Big[
 BW^{\rho_0,\rho_1}((q_2+q_4)^2)(Q+2q_3)(q_2-q_4)+2\Big]
\eea

\noindent
 and 

\bea
&&\kern-20pt BW^{\rho_0,\rho_1}(p^2) = \nonumber \\ 
  &&BW_3(p^2,m_{\rho},\Gamma_{\rho})/m^2_{\rho_0}
  -BW_3(p^2,m_{\rho_1},\Gamma_{\rho_1})/m^2_{\rho_1} \ .\nonumber\\
\eea

For the $\rho-\gamma^*$  coupling we use $g_{\rho\gamma} = 0.1212\ {\rm GeV}^2$.
The double resonant terms disappear in the (-000) and (++--) channels,
 the single resonant contribution, however remains.

For completeness we list all propagators  
required for the current.
A new $\rho_3$ contribution was included in $F_{\rho}\left(Q^2\right)$
 only. The different $\rho$ propagators $T_{\rho}$, $F_{\rho}$ and 
$B_{\rho}$ are used in the current due to the fact that the
$\rho$ may couple in a different way to different resonances and the propagators
themselves contain indirectly some information about the couplings.
\begin{eqnarray}
&&F_{\rho}\left(Q^2,\vec \beta\right)= \frac{1}{1+{\beta}_1+{\beta}_2+{\beta}_3}
 \left[ BW_3(Q^2,m_{\rho},\Gamma_{\rho}) \right.\nonumber\\
&&{\kern +10pt} + {\beta}_1 BW_3(Q^2,\bar m_{\rho_1},\bar{\Gamma}_{\rho_1})
 + {\beta}_2 BW_3(Q^2,\bar m_{\rho_2},\bar{\Gamma}_{\rho_2})\nonumber\\
&&{\kern +10pt}\left. + {\beta}_3 BW_3(Q^2,\bar m_{\rho_3},\bar{\Gamma}_{\rho_3})\right] \ ,
 \label{frhodef}
\end{eqnarray}
where $\vec \beta =( \beta_1,\beta_2,\beta_3 )$ and
\begin{eqnarray}
&&{\kern -30pt} BW_3(Q^2,m_{\rho},\Gamma_{\rho}) =\nonumber \\
&& \frac{m_{\rho}^2} {m_{\rho}^2-Q^2 - i
 \Gamma_{\rho}m_{\rho}\sqrt{\frac{m_{\rho}^2}{Q^2} 
 \left[\frac{Q^2-4m_{\pi}^2}{m_{\rho}^2-4m_{\pi}^2}\right]^3} \ }.
\end{eqnarray}
\noindent
Only the masses $\bar m_{\rho_i}$ and the widths $\bar{\Gamma}_{\rho_i}$
 of $\rho_1,\rho_2$ and $\rho_3$ that appear 
 in $F_{\rho}\left(Q^2,\vec \beta\right)$ were fitted to the data.
The  results are listed in 
 Table \ref{babarfit}.

\begin{table}  [ht]
\vspace{0.5cm}
\begin{tabular}{c|c|c|c|c|c}
 \hline
 $\bar m_{\rho_1}$ & 1.437(2) & $\bar{m}_{\rho_2}$ & 1.738(12)
 &$\bar{m}_{\rho_3}$ & 2.12(2)\\
 \hline
 $\bar{\Gamma}_{\rho_1}$ & 0.520(2) & $\bar{\Gamma}_{\rho_2}$& 0.450(9)
 & $\bar{\Gamma}_{\rho_3}$& 0.30(2)\\
 \hline
 $\beta^{a_1}_1$ & -0.066(3) & $\beta^{a_1}_2$& -0.021(1)
 &$\beta^{a_1}_3$ & -0.0043(5)\\
\hline
 $\beta^{f_0}_1$ & 7(6)$\cdot 10^{4}$ & $\beta^{f_0}_2$& -2.5(5.0) $\cdot 10^{3}$
 &$\beta^{f_0}_3$ & 1.9(1.6) $\cdot 10^{3}$\\
\hline
 $\beta^{\omega}_1$ & -0.33(8) & $\beta^{\omega}_2$& 0.012(3)
 &$\beta^{\omega}_3$ & -0.0053(7)\\
 \hline 
 $c_{a_1}$ & -225(3) & $c_{f_0}$ & 64(3) & $c_{\omega}$& -1.47(4) \\
\hline 
 $c_\rho$ & -2.46(3) & $\chi^2$ & 275 & $n_{d.o.f}$ & 287 \\
  \hline
 \end{tabular}
 \caption{Values of the couplings masses and widths obtained in the fit.
 Masses and widths in GeV; 
 couplings $\beta^j_i$, ($j=a_1,f_0,\omega$ and $i=1,2,3$) as well as
 $c_\rho$ are dimensionless; couplings $c_{a_1}$ and $c_{f_0}$ in GeV$^{-2}$;
  coupling $c_{\omega}$ in in GeV$^{-1}$.}
\label{babarfit}
 \end{table}

For the masses and the widths of particles in all other parts of the current
 we use their PDG values:
\bea
&&m_{\rho} = 0.7755 \ {\rm GeV},\ \ \Gamma_{\rho} = 0.1494\ {\rm GeV}.\nonumber \\
&&m_{\rho_1} = 1.459 \ {\rm GeV},\ \ \Gamma_{\rho_1} = 0.4\ {\rm GeV},\nonumber \\
&&m_{\rho_2} = 1.72 \ {\rm GeV},\ \ \Gamma_{\rho_2} = 0.25\ {\rm GeV}.
\eea

\bea
B_{\rho}\left(Q^2\right) &=&
 \big[ BW_3(Q^2,m_{\rho},\Gamma_{\rho}) \nonumber \\
 &+& \beta BW_3(Q^2,m_{\rho_1}, \Gamma_{\rho_1})\big]/(1+\beta),\\[0.2cm]
&&{\kern -80pt}{\rm with}\ \beta = -0.145\ .
\eea 

The $a_1$ propagator reads:
\bea
 BW_{a_1}\left(Q^2\right) = \frac{m_{a_1}^2}
 {m_{a_1}^2-Q^2 - i \Gamma_{a_1}m_{a_1}\frac{g(Q^2)}{g(m_{a_1}^2)}} \ ,
\eea
\noindent
with \cite{Kuhn:1990ad,Czyz:2000wh}
\bea
 g(Q^2)& =&  1.623 \ Q^2 + 10.38 - \frac{9.32}{Q^2} + \frac{0.65}
        {(Q^2)^2} \nonumber \\
&&{\kern 50pt} {\rm for}  \ \ \ \ Q^2 > (m_{a_1}+m_{\pi})^2,
  \nonumber \\
 g(Q^2) &=&  4.1 \ \left(Q^2-9m_{\pi}^2\right)^3
 \Big[ 1 -3.3 \left(Q^2-9m_{\pi}^2\right) \nonumber \\
&+& 5.8\left(Q^2-9m_{\pi}^2\right)^2
 \Big] \nonumber \\
  &&{\kern+50pt}{\rm for}  \ \ Q^2 < (m_{a_1}+m_{\pi})^2
\eea
\noindent
($Q^2$ in GeV$^2$) and
\beq
m_{a_1}=1.23\ {\rm GeV}, \quad \Gamma_{a_1}=0.2\ {\rm GeV}.
\eeq

\bea
T_{\rho}\left(Q^2\right)&=& 
 \left[ BW_3(Q^2,m_{\rho},\Gamma_{\rho}) 
 + \bar \beta_1 BW_3(Q^2,m_{\rho_1},\Gamma_{\rho_1})\right.\nonumber\\
&+& \left. \bar \beta_2 BW_3(Q^2,m_{\rho_2},\Gamma_{\rho_2})\right]
 /(1+\bar\beta_1+\bar\beta_2),
\eea

\noindent 
where
\beq
\bar \beta_1 = 0.08,\ \ \bar \beta_2 = -0.0075 \ .
\eeq

The $f_0$ meson propagator chosen to be
\beq
BW_{f_0}(Q^2) =
 \frac{m_{f_0}^2} {m_{f_0}^2-Q^2 - i
 \Gamma_{f_0}m_{f_0}\sqrt{\frac{m_{f_0}^2}{Q^2} 
 \frac{Q^2-4m_{\pi}^2}{m_{f_0}^2-4m_{\pi}^2}} \ },
\eeq
\noindent
where
\beq
m_{f_0}=1.35\ {\rm GeV}, \quad \Gamma_{f_0}=0.2\ {\rm GeV}.
\eeq

\bea
\kern-4ptH_{\rho}(Q_1^2,Q_2^2,Q_3^2) &=& BW_3(Q_1^2,m_{\rho},\Gamma_{\rho})
+ BW_3(Q_2^2,m_{\rho},\Gamma_{\rho}) \nonumber \\
\kern-4pt& +& BW_3(Q_3^2,m_{\rho},\Gamma_{\rho}).
\eea

\beq
BW_{\omega}(Q^2) = \frac{m_{\omega}^2}{m_{\omega}^2-Q^2
-im_{\omega}\Gamma_{\omega}}
\eeq
\noindent
is the $\omega$ meson propagator with

\beq
m_{\omega}=0.78265\ {\rm GeV}, \quad \Gamma_{\omega}=0.00849\ {\rm GeV}.
\eeq

To obtain the correct chiral limit \cite{Ecker:2002cw} of the current
\Eq{currentsplit} the following relations should hold

\bea
 c_{a_1} &=& -\frac{4}{3}\frac{1}{f_\pi^2} \nonumber \\
 c_{f_0} + \frac{3}{2}c_{a_1} &=& 4 c_{\rho} \ g_{\rho\pi\pi}^3 \ g_{\rho\gamma}
       \left(\frac{1}{m_{\rho_0}^2} - \frac{1}{m_{\rho_1}^2}\right)^2 \ ,
\label{chiralrel}
\eea
where $f_{\pi} = 0.0924$ GeV.
 Comparing the fitted value
 \bea
 c_{a_1}^{\rm fit} = -225(3)\ \  {\rm GeV}^{-2}
 \eea
 with its
 proper chiral limit
  \bea
 c_{a_1} \simeq -156\ \ {\rm GeV}^{-2} 
 \eea
 and also the second relation from \Eq{chiralrel} as obtained in the fit

  \bea
  c_{f_0}^{\rm fit} + \frac{3}{2}c_{a_1}^{\rm fit} &=& -273(5)\ \ {\rm GeV}^{-2}
 \nonumber \\
  4 c_{\rho}^{\rm fit} \ g_{\rho\pi\pi}^3 \ g_{\rho\gamma}
  \left(\frac{1}{m_{\rho_0}^2} - \frac{1}{m_{\rho_1}^2}\right)^2 &=& - 307(4)\ \ 
  {\rm GeV}^{-2}
  \nonumber \\
\eea

\noindent
  it is clear that they hold only approximately. This reflects the fact that
 the fit was performed in a $Q^2$- region, where the chiral Lagrangian
 leads, at best, to an approximate treatment, and that furthermore higher
 order terms are present.

\bibliography{biblio}

\end{document}